\documentclass[12pt, prd, preprint, amsmath, amssymb, aps, superscriptaddress,eqsecnum, floats,nofootinbib,preprintnumbers]{revtex4-1}

\usepackage{color,graphicx,epsfig}
\usepackage{mathrsfs}
\usepackage{ifpdf}
\usepackage{amsmath}
\usepackage{bm}
\usepackage{color}
\usepackage[dvipsnames]{xcolor}
\usepackage[english]{babel}
\usepackage{graphicx}%
\usepackage{amsfonts}%
\usepackage{amssymb}
\usepackage{braket}
\usepackage{physics}
\usepackage{cancel}
\usepackage{rotating}
\usepackage{array}
\usepackage[colorlinks=true,citecolor=blue,urlcolor=blue,linktocpage=true, linkcolor=blue]{hyperref}
\usepackage{multirow}
\usepackage{amsthm}
\usepackage[nice]{nicefrac}
\usepackage{mathrsfs}
\usepackage{wasysym}
\usepackage{tikz}
\usepackage{subcaption}

\newenvironment{Eqnarray}{\arraycolsep 0.14em\begin{eqnarray}}{\end{eqnarray}}
\def\beqa{\begin{Eqnarray}}
\def\eeqa{\end{Eqnarray}}

\newcommand{\CP}{$\mathcal{CP}$}

\newcommand{\dx}{{\rm d}x}
\newcommand{\dz}{{\rm d}z}
\newcommand{\beq}{\begin{equation}}
\newcommand{\eeq}{\end{equation}}
\newcommand{\bea}{\begin{eqnarray}}
\newcommand{\eea}{\end{eqnarray}}

\newcommand{\be}{\begin{equation}}
\newcommand{\ee}{\end{equation}}
\def\lsim{\mathrel{\rlap{\lower4pt\hbox{\hskip1pt$\sim$}}
    \raise1pt\hbox{$<$}}}         
\def\gsim{\mathrel{\rlap{\lower4pt\hbox{\hskip1pt$\sim$}}
    \raise1pt\hbox{$>$}}}         

\theoremstyle{definition}

\theoremstyle{definition}

\begin{document}
\fontsize{11}{15}\selectfont

\title{Analytic Techniques for Solving the Transport Equations in Electroweak Baryogenesis}

\author{Elina Fuchs}
\email{elinafuchs@uchicago.edu}
\affiliation{Department of Particle Physics and Astrophysics, Weizmann Institute of Science, Rehovot, Israel 7610001}
\affiliation{Fermilab, Theory Department, Batavia, IL 60510, USA}
\affiliation{University of Chicago, Department of Physics, Chicago, IL 60637, USA}

\author{Marta Losada}
\email{marta.losada@nyu.edu}
\affiliation{New York University Abu Dhabi, PO Box 129188, Saadiyat Island, Abu Dhabi, United Arab Emirates}

\author{Yosef Nir}
\email{yosef.nir@weizmann.ac.il}
\affiliation{Department of Particle Physics and Astrophysics, Weizmann Institute of Science, Rehovot, Israel 7610001}

\author{Yehonatan Viernik}
\email{yehonatan.viernik@weizmann.ac.il}
\affiliation{Department of Particle Physics and Astrophysics, Weizmann Institute of Science, Rehovot, Israel 7610001}

\preprint{FERMILAB-PUB-20-202-T, EFI-20-12}

\begin{abstract}
\noindent
\fontsize{12}{13}\selectfont
We develop an efficient method for solving transport equations, particularly in the context of electroweak baryogenesis.
It
provides fully-analytical results under mild approximations and can also test semi-analytical results, which are applicable in more general cases.
Key elements of our method include the reduction of the second-order differential equations to first order, representing the set of coupled equations as a block matrix of the particle densities and their derivatives, identification of zero modes, and block decomposition of the matrix.
We apply our method to calculate the baryon asymmetry of the Universe (BAU) in a Standard Model effective field theory framework of complex Yukawa couplings to determine the sensitivity of the resulting BAU to modifications of various model parameters and rates, and to estimate the effect of the commonly-used thin-wall approximation.
\end{abstract}
\maketitle
\fontsize{11}{9}\selectfont
\tableofcontents
\fontsize{12}{13}\selectfont

\newpage
\section{Introduction}
A long-standing challenge of particle cosmology is to understand the mechanism by which the baryon asymmetry of the Universe (BAU) is generated. The Standard Model (SM) prediction~\cite{Gavela:1993ts,Huet:1994jb} is many orders of magnitude smaller than the observed value of $ Y_B^{\rm obs} \approx 8.6\times 10^{-11} $ measured by PLANCK~\cite{Tanabashi:2018oca}. The requirement for the dynamical process that generates the asymmetry to occur out of thermal equilibrium implies a particular structure for the particle dynamics. In electroweak baryogenesis (EWBG, for reviews, see e.g. Refs.~\cite{Cline:2006ts,Morrissey:2012db,Konstandin:2013caa}), one calculates the asymmetry that is produced during the electroweak phase transition, as bubbles of non-vanishing vacuum expectation value (VEV) of the Higgs field form and expand to fill the Universe \cite{Morrissey:2012db,Cline:2006ts,Joyce:1994bi,Cohen:1994ss,Huet:1995sh,Riotto:1997vy}. The important dynamics in such a scenario arise from the \CP-violating interactions, which occur across the bubble walls and lead to a chiral asymmetry. Weak sphalerons then convert this chiral asymmetry into a baryon asymmetry by acting only on left-handed fermions and changing the baryon number.
The importance of diffusion and the role of leptons was identified in Refs.~\cite{Cohen:1994ss,Cline:2000nw,Chung:2009cb,Guo:2016ixx,deVries:2018tgs,Joyce_1996}. Since the strong sphalerons only wash out the quark asymmetries, and the diffusion into the symmetric phase is larger for leptons, the $\tau$ as a lepton with a sizable Yukawa coupling becomes an efficient source for \CP~violation~\cite{Chung:2009cb,deVries:2018tgs,Fuchs:2020uoc}.

Typically, the calculation is performed semi-classically, such that the particle dynamics is encoded in transport equations - a set of coupled, linear, non-homogeneous differential equations.
The solution to these equations determines the eventual densities of each particle species, yielding a prediction for the baryon asymmetry. The current state-of-the-art approaches for solving these transport equations are the following:
\begin{itemize}
\item Making a set of approximations that simplify the transport equations into a single equation that is analytically solvable and qualitatively understandable~\cite{Chung:2009cb,deVries:2018tgs}; 
\item Solving the full set numerically, which is more accurate but makes it difficult to gain physical insight into the solution \cite{deVries:2018tgs}; 
\item Solving the full set semi-analytically through a process of redefinitions that allow singling out equations to be solved individually as a recursive process \cite{White:2015bva}.
\end{itemize}

We propose a new, semi-analytic method, which is similar to the latter approach,
but simpler in several respects.
Its implementation and usage are clear,
and the understanding of algebraic features provide an intuitive picture of the physical process.
Moreover, under mild approximations, this method allows for a fully-analytic solution, which is useful for estimating the accuracy of the corresponding semi-analytic calculation. Because the approximations are mild, a good agreement between the semi-analytic and the exact solution suggests that the semi-analytic results are reliable also in the original form of the equations and can be extended to more general scenarios.

The paper is organized as follows. In Sections~\ref{sec: con and gen sol} and~\ref{sec: boundary conditions} we solve a general set of transport equations, and impose the suitable boundary conditions. In Section~\ref{sec: zero mode regularization} we discuss the importance of zero modes and illustrate a way to treat them in a numerically stable way. Section~\ref{sec: block decomp} describes techniques that can be applied to produce a fully-analytic solution in applicable cases. In Section \ref{sec: param dep} we apply our method to calculate the baryon asymmetry in several scenarios within the SM effective field theory (SMEFT) framework of complex dimension-six Yukawa terms, testing the sensitivity of the produced asymmetry to modifications by factors of $ \mathcal O (10) $ to model parameters such as the bubble wall parameters and the rates that are an input to the transport equations and have sizable uncertainties. We summarize and discuss our results in Section~\ref{sec: conclusions}. The Appendices \ref{benchmarks} - \ref{app: BAU} provide details of derivations, definitions and benchmark parameters, as well as several consistency checks.

\section{Construction and general solution}\label{sec: con and gen sol}
In the following two sections, we will work in what is known as a two-step approach~\cite{Cline:2000nw,Carena:2002ss}, where the particle dynamics are approximated as a two-step process: In the first step, \CP-violating interactions generate a chiral asymmetry, and in the second step, the weak sphaleron process acts on the chiral density and converts it into a baryon density\footnote{Throughout the paper, the density of a quantity always refers to the \textit{difference} between the corresponding particle and anti-particle densities.}. This decoupling is possible because the weak sphaleron rate is typically slow compared to other processes (see App.~\ref{app:rates}). In App.~\ref{app:12step} we show a comparison between the two-step approach and the one-step approach, where the weak sphaleron is incorporated to the transport equations directly.

The second step consists of solving a single differential equation for the baryon density, and is described in detail in Appendix~\ref{app: BAU}. Solving the transport equations of the first step generalizes the solution of a single equation to a set of equations, one for each particle, and is the focus of this paper.

Taking the diffusion approximation~\cite{Joyce:1994bi,Cohen:1994ss} for the particle density $f$ with the notation
$\partial f \equiv \partial _\mu f^\mu \approx v_wf' - D_ff'' $,
where $v_w$ is the wall velocity and $D_f$ the diffusion coefficient,
a typical set in the two-step approach is the following
\begin{align}\label{transport equations}
\begin{split}
\partial t &= -\Gamma _M^t\mu _M^t - \Gamma _Y^t\mu _Y^t +\Gamma _{ss}\mu _{ss} + S_t\\
\partial b &= -\Gamma _M^b\mu _M^b - \Gamma _Y^b\mu _Y^b +\Gamma _{ss}\mu _{ss} + S_b\\
\partial q &= -\partial t -\partial b\\
\partial \tau &= -\Gamma _M^\tau \mu _M^\tau - \Gamma _Y^\tau \mu _Y^\tau + S_\tau\\
\partial l &= - \partial \tau \\
\partial h &= +\Gamma _Y^t\mu _Y^t - \Gamma _Y^b\mu _Y^b - \Gamma _Y^\tau \mu _Y^\tau \\
\partial u &= +\Gamma _{ss}\mu _{ss}\,.
\end{split}
\end{align}
The \CP-violating sources $S_i$, the $k_i$-functions and the rates $\Gamma_i$ are calculated by standard methods \cite{Cirigliano_2006,Riotto:1997vy,Lee:2004we,deVries:2017ncy} and their values in our framework appear in App.~\ref{benchmarks}. The chemical potentials are related to number densities via $ n_i = T^2 \mu _ik_i/6 + \mathcal{O}(\mu_i^3) $. If we absorb the factor $ T^2/6 $ in the definition of the effective chemical potentials for each process, their values are given by \cite{Trodden:1998ym, deVries:2018tgs}
\begin{align}\label{chemical potentials}
\mu _M^t &= \frac{t}{k_t} - \frac{q}{k_q} \,, & \mu _M^b &= \frac{b}{k_b} - \frac{q}{k_q} \,, & \mu _M^\tau &= \frac{\tau }{k_\tau } - \frac{l}{k_l} \,,\nonumber\\
 \mu _Y^t &= \frac{t}{k_t} - \frac{q}{k_q} - \frac{h}{k_h} \,, & \mu _Y^b &= \frac{b}{k_b} - \frac{q}{k_q} + \frac{h}{k_h} \,, & \mu _Y^\tau &= \frac{\tau }{k_\tau } - \frac{l}{k_l} + \frac{h}{k_h} \,,\\
 &  & \mu _{ss} &= \sum _{i=1}^3\frac{2q_i}{k_{q_i}} - \frac{u_i}{k_{u_i}} - \frac{d_i}{k_{d_i}} \nonumber\,.
\end{align}
The up quark is a representative of the other light quarks ($ d $, $ s $ and $ c $): since they interact only via the strong sphaleron to a good approximation, they are linearly dependent and hence redundant \cite{deVries:2018tgs}.

The sources peak in the broken phase, and for simplicity we approximate the bubble wall as a step function at $z=0$, the center of the bubble wall (see Sec.~\ref{sec:thin wall} for further discussion on this choice). We consider the rates to be constant at each phase (possibly with different values), while for the sources we maintain their $z$-dependence in the broken phase, and eliminate them in the symmetric phase. We thus obtain a set of linear equations with constant coefficients for each phase.
With $N$ denoting the number of species appearing in the transport equations (for the set in Eq.~\eqref{transport equations}, $ N=7 $),
we replace these $N$ equations of second order with $2N$ equations of first order by defining $g_i\equiv f_i'$, such that Eq. \eqref{transport equations} is written in matrix form as
\begin{align*}
\begin{pmatrix}
t' \\ b' \\ \vdots \\ g_t' \\ g_b' \\ \vdots
\end{pmatrix} - \begin{pmatrix}
\ \ \ \ &&&&& \\ & 0_N &&& I_N & \\ && \ \ \ \ &&& \ \ \ \ \\ &&& \frac{v_w}{D_t} && \\ & 0_N &&& \frac{v_w}{D_b} & \\ &&&&& \ddots
\end{pmatrix} \begin{pmatrix}
t \\ b \\ \vdots \\ g_t \\ g_b \\ \vdots
\end{pmatrix} = \begin{pmatrix}
&&&&& \\ & 0_N &&& 0_N & \\ &&&&& \ \ \ \ \\ \frac{\Gamma _t}{D_tk_t} &\cdots&&&& \\ \vdots & \frac{\Gamma _b}{D_bk_b} &&& 0_N & \\ && \ddots &&&
\end{pmatrix} \begin{pmatrix}
t \\ b \\ \vdots \\ g_t \\ g_b \\ \vdots
\end{pmatrix} + \begin{pmatrix}
0\\0\\ \vdots \\ \nicefrac{-S_t}{D_t} \\ \nicefrac{-S_b}{D_b} \\ \vdots
\end{pmatrix}
\end{align*}
\begin{align}\label{trasport - vector notation}
\iff \bar \chi ' - K \bar \chi = \bar S \,, \ \ \ \ K \equiv \begin{pmatrix}
0_N & I_N \\ \Gamma & V
\end{pmatrix} .
\end{align}
Here $\Gamma$ is a matrix of couplings between different particles, where each entry is of the form $\nicefrac{\Gamma _f}{D_f k_f}$. The general solution  to the homogeneous part for each species is a linear combination of modes $f_i(z)=A^{f_i}_j e^{\lambda _j z}$ where $\lambda _j$ are the eigenvalues of $K$. The weights $A^{f_i}_j$ are determined, up to an overall normalization factor, by the eigenvectors of $K$. We can thus write $\bar\chi$ in vector form as follows:
\begin{align}
\bar \chi (z) \equiv \begin{pmatrix}
\bar f(z) \\ \bar g(z)
\end{pmatrix} = \sum _i C_ie^{\lambda _i z} \begin{pmatrix}
\bar f_i \\ \bar g_i
\end{pmatrix} \equiv \hat \Phi(z) \bar C \,,
\end{align}
where $(\bar f_i,\bar g_i)^T$ are the eigenvectors of $K$, and $C_i$ are integration constants. We organize the eigenfunctions in a $ z $-dependent matrix $ \hat \Phi(z) $. Using variation of parameters, the full solution in the broken phase is
\begin{align}\label{nonhomogeneous set solution}
\bar \chi (z) = \hat \Phi (z)\bar C + \hat \Phi (z)\int_0^z \hat \Phi ^{-1}(x)\bar S(x) \dx \,.
\end{align}
We provide the numerical agreement between $g_i$ and $f'_i$ of the solution in App.~\ref{app:derivative}. The impact of including more particles species in the set of transport equations is investigated in App.~\ref{app:nparticles}. Furthermore, in App.~\ref{app:B-L} we show the conservation of $B-L$ numerically.
\section{Boundary conditions}\label{sec: boundary conditions}
In each phase, half of the modes decay and the others diverge or are constant. We choose boundary conditions as follows:
\begin{itemize}
\item In the symmetric phase ($ z<0 $), the integration constants of both the divergent and zero modes are set to 0, complying with the assumption that no baryon asymmetry is present before the electroweak phase transition.
\item In the broken phase ($ z>0 $), the integration constants of divergent modes are used to cancel the divergent integrals coming from the non-homogeneous terms in pairs.
\item The remaining modes are determined by imposing continuity of $ \bar \chi $ at $ z=0 $. Since $ \bar \chi $ contains the vector of derivatives $ \bar g $, this is equivalent to requiring continuity of each particle density and its derivative at $ z=0 $.
\end{itemize}
An important observation is that all modes either decay or are chosen to vanish at infinity, except for the zero modes. These are the only ones to survive deep in the broken phase $ z \to \infty $. Therefore, their existence is crucial for the success of EWBG (and is indeed guaranteed by the linear dependencies in Eq.~\eqref{transport equations}).

The solution of Eq. \eqref{nonhomogeneous set solution} in the broken phase for the $ i $'th component of $ \bar \chi $ is
\begin{align}\label{nonhomogeneous set solution i}
\bar \chi _i(z) &= \hat \phi_{ij} \cdot e^{\hat \lambda_{jk} z} \cdot C_k^B + \hat \phi_{ij} \cdot e^{\lambda _{jk}z} \cdot \hat \phi ^{-1}_{lm}\int_0^z e^{-\lambda _{kl}x} \cdot \bar S_m(x) \dx \,,
\end{align}
where $ \hat \lambda $ is a diagonal matrix constructed from the eigenvalues $ \lambda _j $ and $ \hat \phi _{ij} $ is a matrix of the corresponding eigenvectors.
We denote integration constants of positive (negative) eigenvalues by $ +(-) $, and a $B(S)$ superscript indicates the broken (symmetric) phase. For positive eigenvalues in the broken phase we choose
\begin{align}\label{Cj for positive lambda}
C_k^{B+} = -\hat \phi _{lm} ^{-1} \int_0^\infty e^{-\lambda _{kl}x} \cdot \bar S_m(x) \dx \,.
\end{align}
This choice guarantees convergence at infinity. The continuity conditions are treated as follows. In the broken phase at the phase boundary, Eq.~\eqref{nonhomogeneous set solution i} reads
\begin{align}\label{phi_ij C_j = b_i}
\bar \chi (z \to 0^+) = \hat \phi _{ij}^BC^B_j = \hat \phi _{ij}^BC^{B+}_j + \hat \phi _{ij}^BC^{B-}_j \equiv b_i + \hat \phi _{ij}^BC^{B-}_j ,
\end{align}
where $\bar b$ is a constant vector with entries $b_i$ obtained from Eq. \eqref{Cj for positive lambda}.

In the symmetric phase, we set the integration constants associated with negative eigenvalues to zero, such that
\begin{align}
\bar \chi (z \to 0^-) = \hat \phi _{ij}^SC^{S+}_j + 0 \,.
\end{align}
Continuity at $ z=0 $ is then
\begin{align}\label{c - c = b}
\hat \phi _{ij}^SC^{S+}_j - \hat \phi _{ik}^BC^{B-}_k &= b_i \,, & 1\leqslant j \leqslant N ,\;\; N+1 \leqslant k \leqslant 2N ,\;\; 1 \leqslant i \leqslant 2N\,.
\end{align}
To reach the final expressions, we need to solve a linear set of equations for the remaining integration constants. We can organize these constants in a vector $ \bar c \equiv (C^{S+} , C^{B-})^T $ and the corresponding modes as columns of a matrix $ \hat A \equiv \left( \phi^S | -\phi^B \right) $, such that finding the remaining integration constants $ \bar c $ amounts to solving the equation $ \hat A \bar c = \bar b $. We can then collect the relevant densities, which in the two-step approach involves summing over the densities of the left-handed multiplets in the symmetric phase. In the case of Eq.~\eqref{transport equations}, we recall that $ u $ acts as a representative of the light quarks. To obtain the densities of the left-handed multiplets of the first two generations, we relate them to $ u $ via $ q_1 = q_2 = -2u $ \cite{deVries:2018tgs}. The chiral density is $ n_L = q + l - 4u $, which we plug into Eq.~\eqref{Y_B} to solve for the baryon asymmetry. Note that since we only need the zero modes for our final result of $Y_B$, Eqs.~\eqref{Cj for positive lambda} of $\lambda_j=0$,~\eqref{c - c = b} and \eqref{Y_B} imply that $ Y_B $ is exactly linear in the integrated \CP-violating sources $ S_f $.

\subsection{One step and two step approaches}
To obtain the baryon asymmetry in the one-step approach, we need to add to Eq. \eqref{transport equations} the following terms:
\begin{align*}
\partial q &\to \partial q - 3\Gamma _{\text{ws}}\mu _{\text{ws}} \,, & \partial l &\to \partial l - \Gamma _{\text{ws}}\mu _{\text{ws}} \,, & \mu _{\text{ws}} &= \sum _i \frac{l_i}{k_{l_i}} + 3\frac{q_i}{k_{q_i}} \,.
\end{align*}
In this case, the degeneracy among light quarks in Eq. \eqref{transport equations} is explicitly broken. Therefore we must reintroduce at least one left-handed quark multiplet. We may keep one quark generation implicit as long as we add its contribution to $ Y_B $ in the end. The baryon density is obtained by summing over the zero modes of each species in the broken phase, and multiplying the quark densities by $ 1/3 $. The convergence of the two-step approach towards the one-step solution for small $\Gamma_\text{ws}$ is shown in App.~\ref{app:12step}.

\section{Zero modes and numerical regularization}\label{sec: zero mode regularization}
We have seen that zero modes are crucial for the generation of a baryon asymmetry, since the rest of the modes necessarily decay deep within the broken phase. Here we show explicitly that the existence of the zero modes is guaranteed by the structure of the transport equations, and then discuss their impact on the numerical analysis. Consider again the matrix $K$ in Eq.~\eqref{trasport - vector notation}. Zero is an eigenvalue of $ K $ iff $\det K = 0$. The determinant of a block matrix
$M= \begin{pmatrix}
     A & B\\ C & D
    \end{pmatrix}
$
for invertible $D$ is $\det(M)= \det(A - BD^{-1}C) \det(D)$. With $A=0, \ B=I_N, \ C = \Gamma $, we obtain
\begin{align*}
\det(K)= \det(-\Gamma ) = (-1)^N\det(\Gamma).
\end{align*}
The block $\Gamma$ corresponds to couplings in the transport equations, which we know are not all linearly independent: In the two-step approach, the couplings of left-handed multiplets are the negatives of the corresponding right-handed ones (e.g. $ \partial l = -\partial \tau $). Thus each generation produces a zero mode. In the one-step scenario, the degeneracy is broken between left and right, but reintroduced across species. For example,
\begin{align*}
\begin{cases}
\partial q = -\partial t -\partial b - 3\Gamma _{\text{ws}}\mu _{\text{ws}}\\
\partial l = -\partial \tau - \Gamma _{\text{ws}}\mu _{\text{ws}}
\end{cases} \Rightarrow \partial q = -\partial t -\partial b + 3\partial l + 3\partial \tau \,.
\end{align*}

When incorporating many particle species in the transport equations, finding the eigenvalues of $K$ is an intrinsically numerical task, equivalent to finding roots of high-order polynomials. The zero modes, which necessarily exist, may cause numerical instabilities if not treated carefully. A way to circumvent the problem is to first perform a partial diagonalization of $K$ to extract the zero eigenvalues, and then solve for the rest of the system independently. Let us outline the procedure. Suppose we have a matrix $M$ for which we know only a subset $j$ of its $m$ eigenvalues. We would like to find a matrix $ U $ such that
\begin{align}
M' = U^{-1}MU = \begin{pmatrix}
\tilde D & 0 \\ 0 & \tilde M
\end{pmatrix} ,
\end{align}
where $\tilde D$ is diagonal and consists of the $ j $ known eigenvalues of $M$, and $\tilde M$ is arbitrary. If $ M $ is diagonalizable, then in particular it is partially-diagonalizable. In the case of Eq. \eqref{transport equations}, $ \tilde D = 0_{2\times 2} $. If we only diagonalize a block of $ M $, then we have $ U = \left( U_r | V_r \right) $ where $ U_r $ consists of the right-eigenvectors that were already found, and $ V_r $ remains to be determined. We can write $ U^{-1} = \left( \frac{U_l}{V_l} \right) $ where $ U_l $ are the left-eigenvectors and $ V_l $ the remainder. We have
\begin{align}
I_N = U^{-1}U = \left( \frac{U_l}{V_l} \right)\left( U_r|V_r \right) \iff \begin{cases}
U_l U_r = I_j \,,\\
V_l V_r = I_{m-j} \,, \\
U_l V_r = V_l U_r = 0 \,.
\end{cases}
\end{align}
From this, we see that $ V_r $ and $ U_l $ span orthogonal spaces, such that
\begin{align}
U = \left( U_r | \left( U_l^t \right)^\perp \right).
\end{align}
The upshot in our case is that we found a way to reduce the original problem of finding the eigenvalues of the \textit{singular} matrix $ K $ to finding the eigenvalues of a \textit{regular} matrix $ \tilde K $, which should be numerically stable. Going back to the general case, we now need to match the eigensystem of the transformed matrix $ M' $ to that of the original matrix $ M $. The eigenvalues are the same, as can be seen from
\begin{align*}
\det(U^{-1}MU-\lambda I) = \det(U^{-1}\left( M-\lambda I \right)U) = \det U^{-1} \det U \det(M-\lambda I) = \det(M-\lambda I) \,.
\end{align*}
For the eigenvectors, suppose $ y $ is an eigenvector of the transformed matrix, and denote $ x = Uy $. Then,
\begin{align*}
U^{-1}MUy = \lambda y \iff U^{-1}MU U^{-1}x = \lambda U^{-1}x \iff Mx = \lambda x \,.
\end{align*}
We find that if $ y $ is an eigenvector of $ M' $ corresponding to an eigenvalue $ \lambda $, then $ x $ is an eigenvector of $ M $, corresponding to the same eigenvalue $ \lambda $.

To summarize the procedure, we start by finding the eigenvectors of the zeros of $ K $ to obtain $ U $, which we use to partially diagonalize $ K $. We then find the eigensystem of $ K' $, and transform the eigenvectors to obtain the eigensystem of the original matrix.

\section{Block decomposition for analytical solution}\label{sec: block decomp}
In this section we show that, under certain approximations, we can obtain a fully analytic solution. This is useful for checking the semi-analytic method, where the eigenvalue problem is solved numerically, and consequently all downstream calculations are numeric as well. Since the approximations we are going to use are mild, finding that the results are in good agreement means we should expect the semi-analytic method to be reliable also for the exact equations.

Using the general structure of $ K $, we get
\begin{align}
K \bar \Phi ^i = \lambda _i \bar \Phi ^i \iff \begin{pmatrix}
0 & I_N \\ \Gamma & V
\end{pmatrix} \begin{pmatrix}
\bar \phi ^i \\ \bar \varphi ^i
\end{pmatrix} = \begin{pmatrix}
\lambda _i\bar \phi ^i \\ \lambda _i\bar \varphi ^i
\end{pmatrix}
\end{align}
\begin{align}\label{varphi = lambda phi}
\Rightarrow \begin{cases}
\bar \varphi ^i = \lambda _i \bar \phi ^i\\
\Gamma \bar \phi ^i + V \bar \varphi ^i = \lambda \bar \varphi ^i
\end{cases} \Rightarrow \Gamma \bar \phi ^i + \lambda V \bar \phi ^i = \lambda^2 \bar \phi ^i \,.
\end{align}
We obtained equations that are, first, independent of $\varphi^i$, and second, close to representing an eigenvalue problem for an $ N\times N $ matrix instead of $ 2N\times 2N $. If we assume the diffusion coefficients are all the same, then $ V $ becomes a scalar matrix, and we obtain an actual eigenvalue problem for the matrix $ \Gamma $, with eigenvalues
\begin{align}\label{lambda tilde}
\tilde \lambda \equiv \lambda ^2-\lambda V \,,
\end{align}
given by
\begin{align}\label{eigensystem lambda tilde}
\Gamma \bar \phi ^i = \tilde{\lambda }\bar \phi ^i \,.
\end{align}
This can also be seen from determinant properties of block matrices. For a general matrix, if $ A $ is invertible, then
\begin{align*}
\det\begin{pmatrix}
A & B \\ C & D
\end{pmatrix} = \det(A) \det (D - CA^{-1}B) \,.
\end{align*}
In our case,
\begin{align*}
A &= \begin{pmatrix}
-\lambda && \\ & \ddots & \\ && -\lambda
\end{pmatrix} , &
B &= I_N \,, &
C &= \Gamma \,, &
D &= \begin{pmatrix}
V_1-\lambda && \\ & \ddots & \\ && V_N-\lambda
\end{pmatrix} ,
\end{align*}
such that if all the diffusion coefficients are the same (i.e. $ \forall i,j : V_i = V_j \equiv V $), we get for the non-zero eigenvalues
\begin{align}\label{det(M)}
\det (K-\lambda ) &= (-\lambda )^N \det(V- \lambda + \frac{1}{\lambda }\Gamma ) \nonumber\\
&= \det((\lambda ^2 - V\lambda) - \Gamma ) \equiv \det(\tilde{\lambda } - \Gamma ) \,.
\end{align}
Neglecting the Higgs density $ h \approx 0 $ and decoupling the weak sphaleron (two-step approach) allows us to solve the eigenvalue problem \eqref{eigensystem lambda tilde} fully analytically. Doing so allows us to obtain the eigenvectors of $ K $ by solving the quadratic equations \eqref{lambda tilde} for $ \lambda _i^\pm $ and using the relation \eqref{varphi = lambda phi} to construct $ \bar \Phi ^i $.

We can obtain an analytic solution also without assuming the diffusion coefficients are all equal, and instead approximate them as equal only among fields from the same family:
\begin{align}
\begin{cases}
D_q = D_t = D_b = 6/T \,,\\
380/T = D_\tau \approx D_l = 100/T \,.
\end{cases}
\end{align}
This approximation allows us to arrange $ \Gamma $ in blocks of equal $ D $'s for quarks and leptons separately. Then, each block is a subproblem of the original eigenvalue problem, which will be solved separately. If we look at Eq.~\eqref{transport equations} under the above approximations, we have $ q,t,b,u,l,\tau $, which naively form a $ 12\times 12 $ matrix, but reduces to separate $ 4\times 4 $ and $ 2\times 2 $ blocks, which are easily solvable. Of course, this decomposition works also under the more aggressive approximation of equal diffusion coefficients. If we do not make the approximation $ D_l \approx D_\tau $, then the quark block still forms an eigenvalue problem with an effective eigenvalue $ \tilde \lambda _q \equiv \lambda ^2 - \lambda V_q $, but the lepton block does not. Instead, it is just a set of two equations in 3 variables: $ \lambda , \phi _l , \phi _\tau $, where we denote for simplicity the latter two by $ l,\tau $, respectively. The equations are thus
\begin{align}
\begin{cases}
\Gamma _{ll} l + \Gamma _{l\tau} \tau + \lambda V_l l - \lambda ^2 l = 0 \,,\\
\Gamma _{\tau l} l + \Gamma _{\tau \tau} \tau + \lambda V_\tau \tau - \lambda ^2 \tau = 0 \,.
\end{cases}
\end{align}
Setting $ \tau = 0 $ immediately implies $ l = 0 $, trivializing the solution. We can therefore choose $ \tau = 1 $, which gives
\begin{align}
l = \frac{1}{\Gamma _{\tau l}}\left( \lambda ^2 - \lambda V_\tau - \Gamma _{\tau \tau}  \right) ,
\end{align}
\begin{align}
\Rightarrow \left( \Gamma _{ll} + V_l \lambda - \lambda ^2 \right)\left( \lambda ^2 - V_\tau \lambda - \Gamma _{\tau \tau } \right) + \Gamma _{l\tau }\Gamma _{\tau l} = 0 \,.
\end{align}
Plugging in the values of $ \Gamma _{ij} $ eliminates the constant term, reproducing the expected zero eigenvalue, and leaving us with $ \left( \Gamma \equiv \Gamma _M + \Gamma _Y \right) $
\begin{align}
\lambda ^3 - \left( \frac{v_w}{D_\tau } + \frac{v_w}{D_l} \right)\lambda ^2 + \left( \frac{v_w^2}{D_l D_\tau } - \frac{\Gamma }{k_l D_l} - \frac{\Gamma }{k_\tau D_\tau } \right)\lambda + \frac{v_w \Gamma }{D_l k_\tau D_\tau } + \frac{v_w \Gamma }{D_\tau k_l D_l} = 0 \,.
\end{align}
Note that the zero eigenvalue determines the eigenvector to be
\begin{align}
(l,\tau ) = \left( \frac{k_l}{k_\tau } , 1 \right) .
\end{align}
The other eigenvalues are given numerically by
\begin{align}
\lambda \in \{0.044, -0.067,0.11\} \,,
\end{align}
where $ \lambda = 0.044 $ leads again to the eigenvector
\begin{align}
(l,\tau ) = \left( \frac{k_l}{k_\tau } , 1 \right) ,
\end{align}
and the other two eigenvalues both produce the eigenvector
\begin{align}
(l,\tau ) = \left( -1, 1 \right) .
\end{align}
The apparent degeneracy in the eigenvectors is resolved when we construct the \textit{full} eigenvectors using $ \bar \varphi ^i = \lambda _i \bar \phi ^i $, where all the quark entries are 0.

The next step in the process is to invert the matrix $ \hat \Phi $ corresponding to the eigenvectors of $K$. Instead of directly inverting $ \hat \Phi $, which is computationally taxing, we will follow a similar path to the regularization procedure, and find the eigenvectors of $ K^T $ as we did for $ K $. These will be the left eigenvectors of $ K $, and when properly normalized construct the inverse of $ \hat \Phi $. Denote the left eigenvectors by $ (\bar u ,\bar v) $, such that
\begin{align}
\begin{pmatrix}
0 & I_N \\ \Gamma & V
\end{pmatrix}^T \begin{pmatrix}
\bar u \\ \bar v
\end{pmatrix} = \begin{pmatrix}
0 & \Gamma ^T \\ I_N & V
\end{pmatrix} \begin{pmatrix}
\bar u \\ \bar v
\end{pmatrix} = \lambda \begin{pmatrix}
\bar u \\ \bar v
\end{pmatrix}
\end{align}
\begin{align}
\Rightarrow \begin{cases}
\Gamma ^T \bar v = \lambda \bar u \\
\bar u + V \bar v = \lambda \bar v
\end{cases} \Rightarrow
\bar u = (\lambda - V) \bar v \,, \ \ \ \Gamma ^T \bar v = (\lambda ^2 - \lambda V) \bar v \,.
\end{align}
The effective eigenvalues are again the same, and the eigenvectors $ \bar v^i $ are solved for and used to obtain $ \bar u^i $. We organize the left eigenvectors as rows in a matrix $ A $, and choose their normalization such that $ A = \hat \Phi ^{-1} $. From here on we simply follow with the semi-analytic procedure, and eventually plug in the numbers for the baryon asymmetry with arbitrary precision.
The agreement between the fully and the semi-analytical solution is numerically investigated for two different sets of assumptions in App.~\ref{app:semi-analytic}.

\section{Parameter dependence}\label{sec: param dep}
In this section we discuss how various model parameters and rates affect the baryon asymmetry. Our detailed calculations are performed in the framework of a Standard Model Effective Field Theory (SMEFT) with dimension-six complex Yukawa terms. This framework thus introduces new sources of \CP~violation, but does not enhance the electroweak phase transition which is assumed to be addressed separately.
The Lagrangian for dimension 4 and dimension 6 Yukawa-type terms is given by:
\beq\label{eq:lagd4}
{\cal L}_{\rm Yuk}= y_f\left(\overline{F_L}F_RH+
\frac{2}{v_0^2}(T_R^f+iT_I^f)|H|^2\overline{F_L}F_RH \right)+ {\rm h.c.,}
\eeq
where $ v_0 $ is the Higgs VEV at zero temperature. 
The definitions of relevant quantities and the benchmark values for the numerical calculations are given in Appendix \ref{benchmarks}. In particular, the benchmark values for $ T_I^f $ appear in Table~\ref{tab:rates T=0}. In the examples shown here, we set $ T_R^f = 0 $ for all species. 
The phenomenology of the muon and third-generation fermions, including the interplay of $T_R^f$ and $T_I^f$, is analyzed in detail in Refs.~\cite{Fuchs:2019ore,Fuchs:2020uoc}.
For $ T_R^f = 0 $, see also Refs.~\cite{deVries:2017ncy,deVries:2018tgs}.

\subsection{Relaxation and Yukawa rates}
\begin{figure}[h]
\centering
\includegraphics[width=0.85\linewidth]{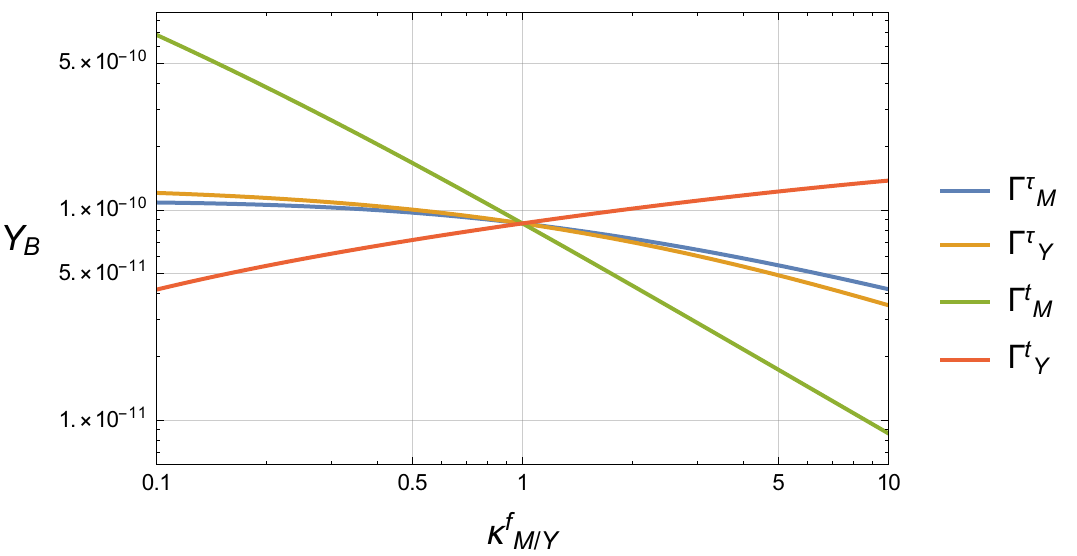}
\caption{$ Y_B $ as a function of the modifier $ \kappa_{M/Y}^f $ of the relaxation and Yukawa rates
shown for the $\tau$ and $t$ sources.
}
\label{fig:gammap}
\end{figure}

Consider the relaxation rates $ \Gamma _M,\Gamma _Y $ that appear in Eq.(\ref{transport equations}) and explicitly defined in Eq.~(\ref{Gamma M,Y expression}). These are \CP-conserving terms that for large values tend to produce chemical equilibrium and dampen the asymmetry. They are calculated to leading order in perturbation theory. Higher-order corrections and terms beyond the underlying approximations are expected to modify these rates, see {\it e.g.} Ref.~\cite{Lee:2004we,Postma:2019scv}, and consequently have an impact on the calculated baryon asymmetry. Here we do not include these higher-order terms. Instead, we study the sensitivity of the baryon asymmetry to modifications of $\Gamma_M$ and $\Gamma_Y$. In Figure~\ref{fig:gammap} we replace
\begin{align}
\Gamma _{M/Y}^f \to \kappa_{M/Y}^f\Gamma _{M/Y}^f \,, \label{eq:kappaMY}
\end{align}
and plot $ Y_B $ as a function of the modifiers $ \kappa_{M/Y}^f$, allowing for large deviations from the leading-order value.

\begin{figure}[h]
\centering
\begin{subfigure}{.48\textwidth}
  \centering
  \includegraphics[width=1\linewidth]{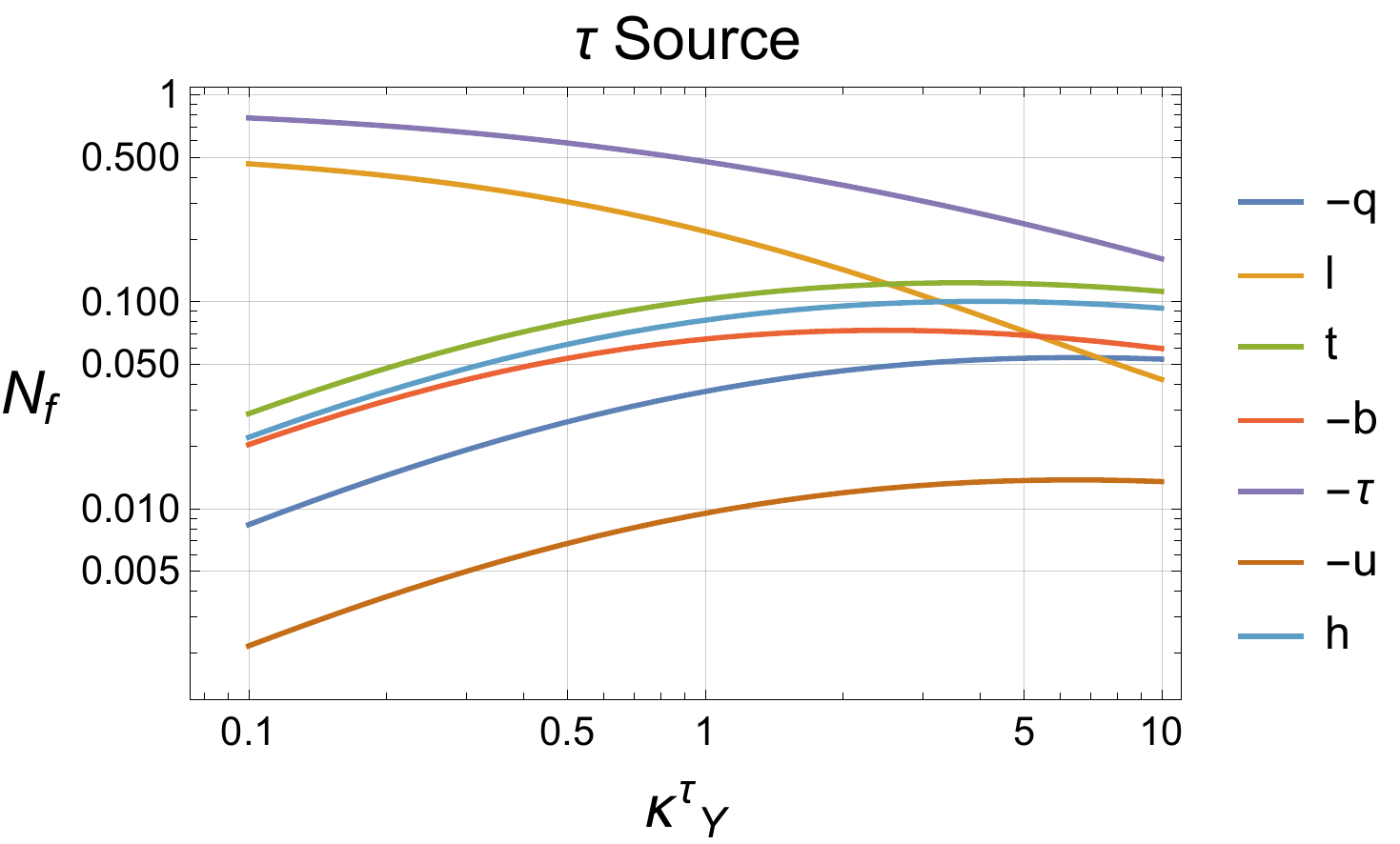}
\end{subfigure} \ \ \ \
\begin{subfigure}{.48\textwidth}
  \centering
  \includegraphics[width=1\linewidth]{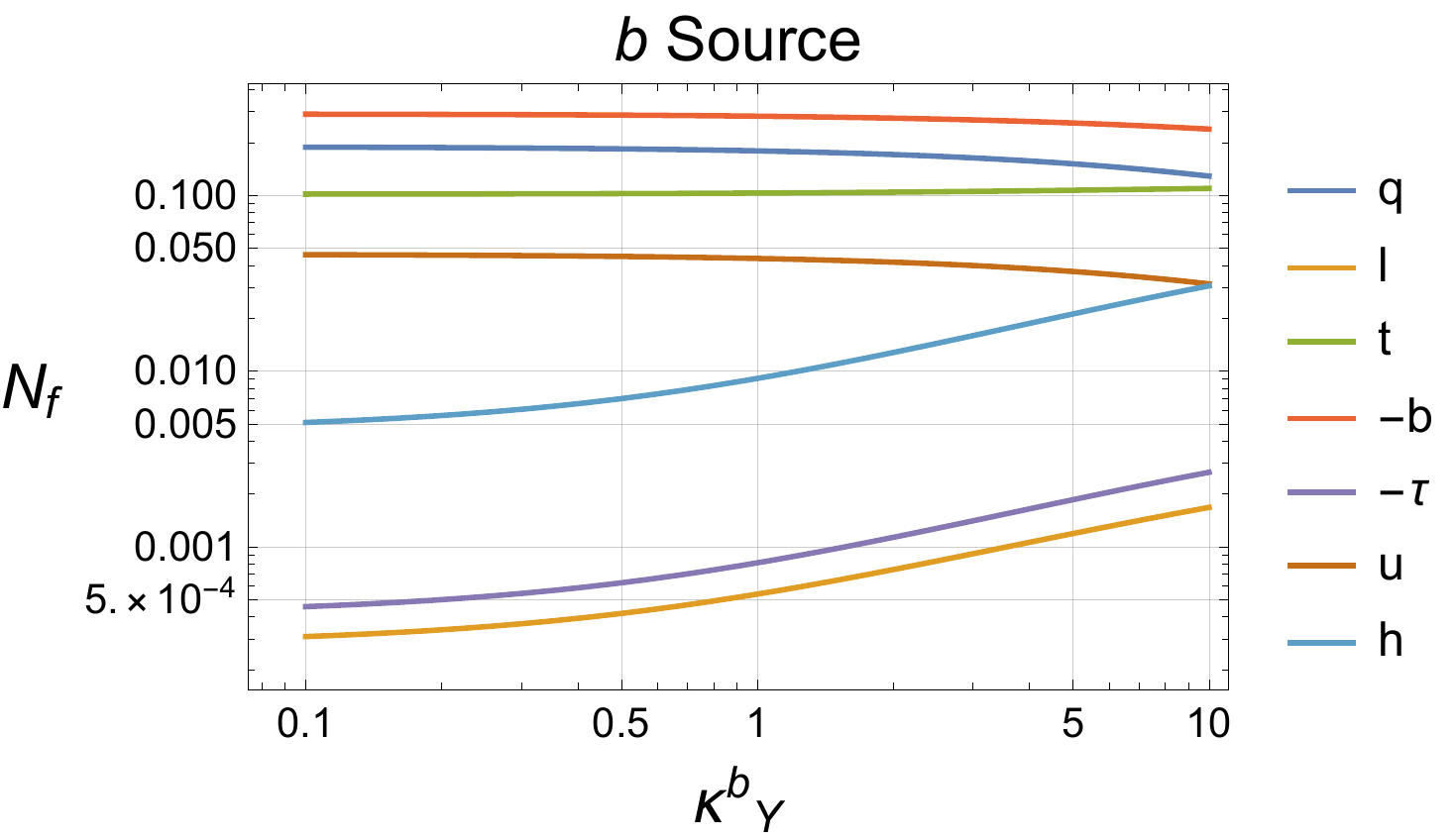}
\end{subfigure}\newline \\
\begin{subfigure}{.48\textwidth}
  \centering
  \includegraphics[width=1\linewidth]{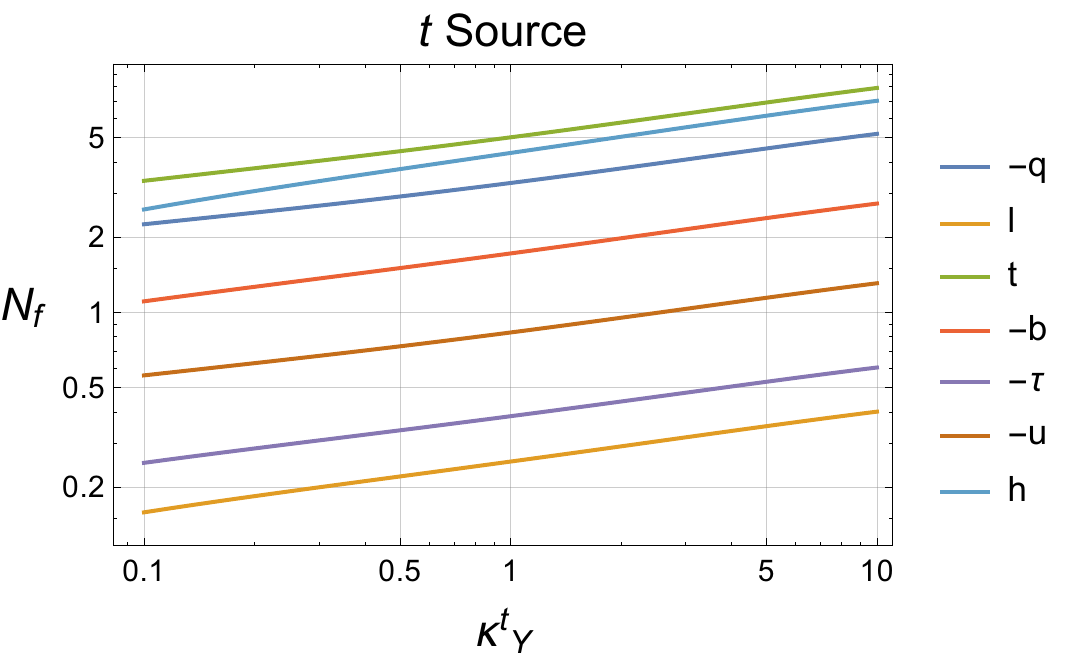}
\end{subfigure}
\caption{Particle densities integrated in the symmetric phase $ N_f $ prior to weak sphaleron action as a function of the Yukawa rate modifier $ \kappa_Y^f$. In each subfigure, a different source is turned on, and the corresponding fermion's Yukawa rate is modified.
}
\label{fig:bysph1densp}
\end{figure}

\begin{figure}[h]
\centering
\begin{subfigure}{.49\textwidth}
  \centering
  \includegraphics[width=1\linewidth]{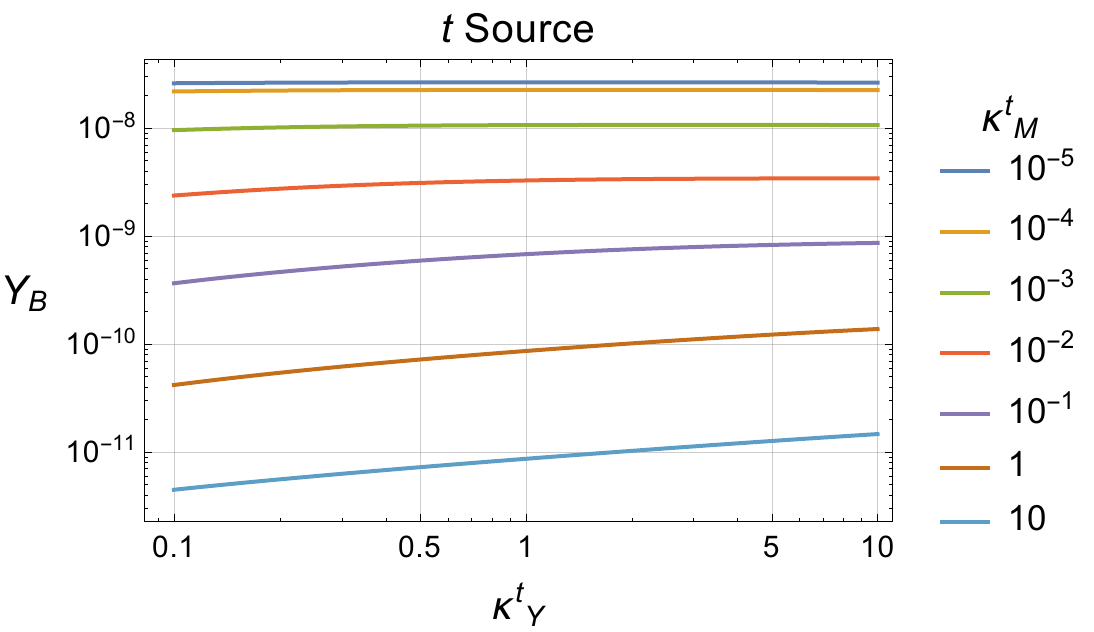}
\end{subfigure} \ \ \ \
\begin{subfigure}{.46\textwidth}
  \centering
  \includegraphics[width=1\linewidth]{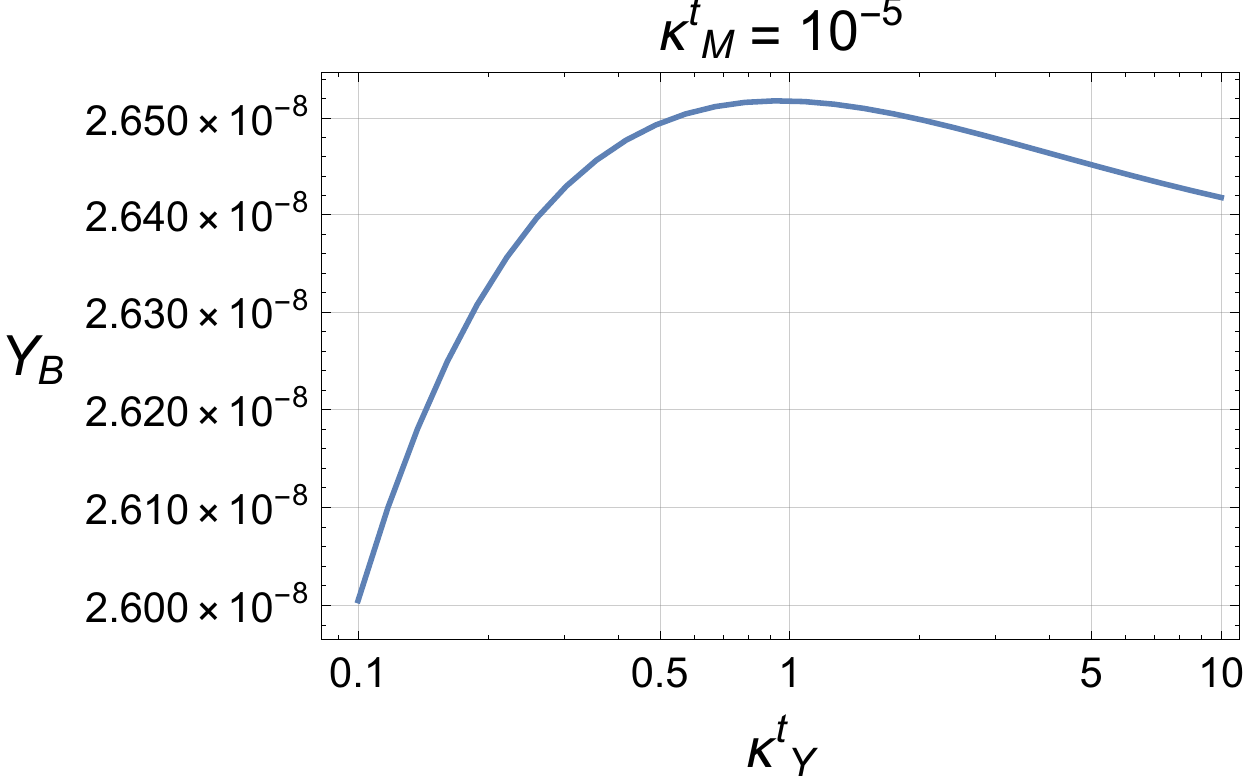}
\end{subfigure}
\caption{\textit{Left}: $ Y_B $ as a function of $ \kappa_Y^t $ for a top source. Each curve is for a particular value of the relaxation rate, specified by $ \kappa_M^t $.
\textit{Right}: A zoomed-in version on the curve $ \kappa_M^t  = 10^{-5} $, which corresponds to $ \Gamma_M^t \approx 0.01 \mathrm{GeV} \approx \frac{1}{5}\Gamma _M^b$.
}
\label{fig:tymp}
\end{figure}

For the tau, changes of $\mathcal O(10)$ to the rates translate to only $\mathcal O(1)$ changes in $Y_B$. The top is much more sensitive to changes in the relaxation rate due to its large mass: an $ \mathcal O(10) $ increase (decrease) of $ \Gamma _M $ produces an $ \mathcal O(10) $ decrease (increase) in $ Y_B $. On the other hand, the larger $\Gamma_Y$, the larger $ Y_B $. This may be an effect of avoiding the washout due to $ \Gamma _M $ by transferring some density to other species with slower rates. To illustrate this point further, we integrate the number densities of each particle species in the symmetric phase, prior to the weak sphaleron action. We denote the integrated density of particle $ f $ in the symmetric phase by
$ N_f = \int_{-\infty}^0 {\rm d}z~n_f(z)$.
In Fig.~\ref{fig:bysph1densp}, we show for each source how the integrated densities are affected by modification to the Yukawa rate.
For a $ \tau $ source, we see that the densities for $ \tau ,l $ (right-handed tau and left-handed third generation lepton doublet) are mostly dominant, but decrease as the the Yukawa rate for the tau is increased, while other particle species increase in density. For a $ b $ source, it is $ b,q $ (right-handed bottom and left-handed third generation quark doublet) which are dominant, again showing a mild increase in other particle densities at their own expense as $ \Gamma _Y^b $ increases. We also have a slight decrease in the density of $ u $, the representative of the light quarks, as these get sourced predominantly by the strong sphaleron, considering the smallness of their Yukawa couplings. Thus a decrease in the bottom density results in less chemical potential for strong sphaleron interactions and less accumulation of light quarks. Finally, for a $ t $ source, we see an increase in the density of every particle species. Interestingly, it is not the left-handed quark doublet that contributes most to the baryon asymmetry via the weak sphaleron, because the strong sphaleron quickly spreads the quark density among the quarks, and $ q $ is almost canceled against $ q_2 + q_1 = -4u $. Rather, it is the left-handed leptons, enhanced by large Yukawa interactions of the top, that drive the weak sphaleron into increasing the baryon asymmetry. The reason all densities increase in the top case is that the top relaxation rate is the strongest source of washout, and we see here that by increasing the Yukawa rate, all other species, which experience much less washout, increase in density. To show that the relaxation rate of the top is responsible for this behavior, we show in Fig.~\ref{fig:tymp} the effect of changing the Yukawa rate for various values of the relaxation rate.
We see that for large values of $ \Gamma _M $, there is a positive effect of $ \Gamma ^Y $ on $ Y_B $. For small $ \Gamma _M $, the effect decreases, and in the order of $ \Gamma _M^t \sim \Gamma _M^b $, the slope vanishes, and an opposite trend emerges (albeit with a diminished amplitude).

We also note that turning off the Yukawa rate in the symmetric phase and neglecting the Higgs density reverses this behavior, as well as flips the overall chiral excess and hence the produced baryon asymmetry. In this case, we would require a CPV operator with a coefficient of opposite sign. This emphasizes the impact of the kinetic redistribution of densities that occurs in the transport equations. In Table~\ref{tab:rate mod} we provide a summary of the effects seen in Fig.~\ref{fig:gammap} for such typical modifications that may occur given more precise calculation of the relaxation rates.

\setlength{\tabcolsep}{10pt}
\renewcommand{\arraystretch}{0.9}
\begin{table}[h]
\centering
\begin{tabular}{c|l l|l l}
\hline \hline
Particle & $ 0.1\Gamma_M^B $ & $ 10\Gamma_M^B $ & $ 0.1\Gamma_Y $ & $ 10\Gamma_Y $ \\
\hline
$ \tau $ & 1.3 & 0.5 & 1.4 & 0.4 \\
$ \mu  $ & 1.009 & 0.93 & 1.008 & 0.93 \\
$ t $ & 7.9 & 0.1 & 0.5 & 1.6 \\
$ b $ & 1.1 & 0.7 & 0.99 & 1.0004 \\
\hline \hline
\end{tabular}
\caption{The ratio $ Y_B^{\text{mod}}/Y_B $ of the modified to unmodified predictions of the baryon asymmetry, for particular values of the modifier $ \kappa_{M/Y} $. Every modification is made only with the corresponding active source term. Effects of modifications to relaxation rates of species with no active source is smaller than for the particle with the active source.}
\label{tab:rate mod}
\end{table}

\subsection{Sphaleron rates}\label{sec: sphaleron rates}
\begin{figure}[h]
\begin{subfigure}{.48\textwidth}
  \centering
  \includegraphics[width=1\linewidth]{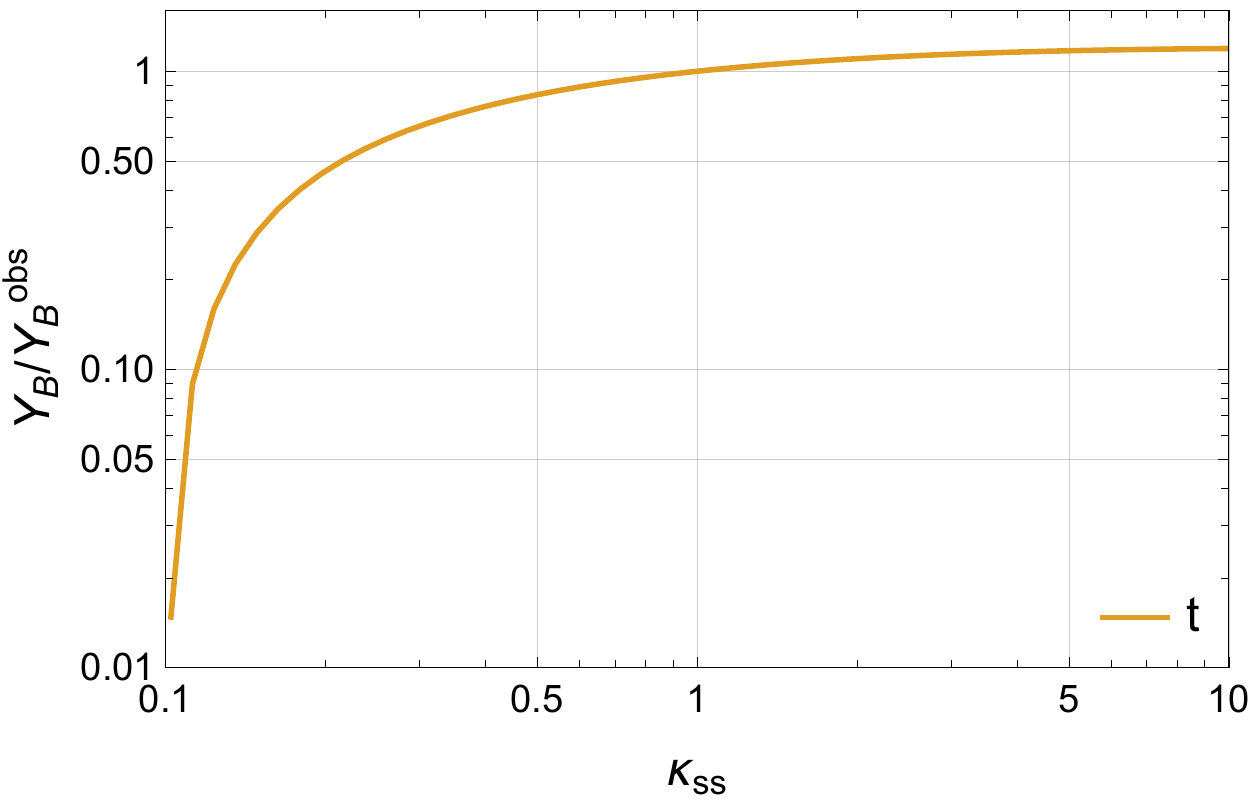}
\end{subfigure}\ \ \ \
\begin{subfigure}{.48\textwidth}
  \centering
  \includegraphics[width=1\linewidth]{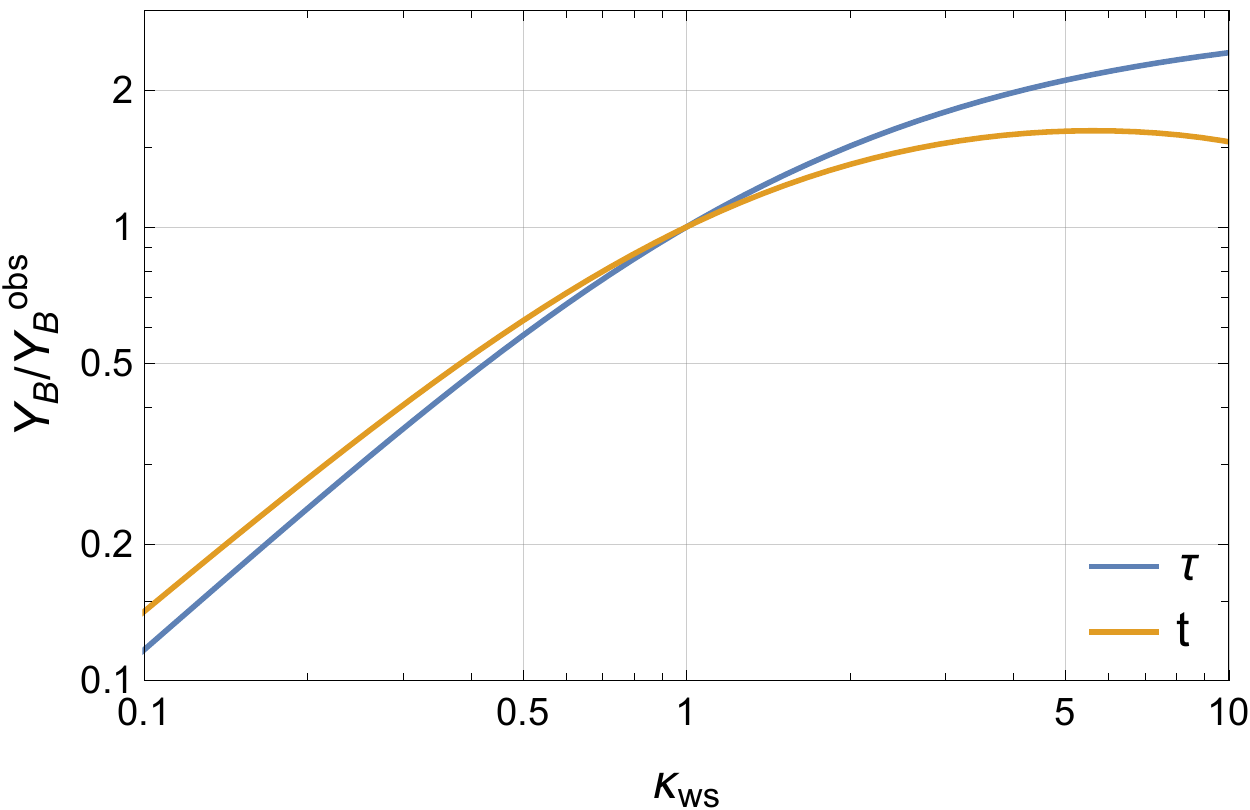}
\end{subfigure}
\caption{$ Y_B/Y_B^{\rm obs}$ as a function of the strong (left) and weak (right) sphaleron rates modifier $\kappa_{\rm ss}, \kappa_{\rm ws} $. For the strong sphaleron, we show only the top source, as the tau source is virtually unaffected by such modifications.}
\label{fig:sph rates}
\end{figure}

The sphaleron rates are similarly subject to uncertainties~\cite{Moore:1997im,Moore:2000mx,DOnofrio:2014rug}. It is interesting to compare the sensitivities to these parameters between the case of a $ t $ source and a $ \tau $ source. Introducing similar modifiers, $ \kappa _{\text{ss}} $ and $ \kappa _{\text{ws}} $, Fig.~\ref{fig:sph rates} shows that the top-sourced BAU is suppressed when the strong sphaleron rate is decreased. The tau, in comparison, is virtually unaffected by modifications to the strong sphaleron rate: an $ O(10) $ modification to $ \Gamma _{\text{ss}} $ with a tau source changes $ Y_B $ by about about $ 0.1\% $ (not shown in the figure). This is because the strong sphaleron acts solely on quarks, which are only weakly coupled to the lepton sector via the Higgs, and therefore have little impact in the case of a lepton source.
Changes in the weak sphaleron rate impact the baryon asymmetry similarly for both $ \tau $ and $ t $, as seen in the right plot of Fig.~\ref{fig:sph rates}.

\subsection{Ultra-thin wall approximation}\label{sec:thin wall}
Approximating the relaxation rate $ \Gamma _M $ as a step function requires choosing the point where it is turned on/off, which is essentially choosing the position of the bubble-wall. This is the ultra-thin wall approximation, and is a necessary step in the matrix formalism (see \cite{White:2015bva,deVries:2017ncy}, and also \cite{deVries:2018tgs} for a direct comparison between the characteristic bubble wall width $ L_w $ and other typical length scales). This choice is somewhat arbitrary, since the actual bubble-wall has a smooth profile characterized by $ \phi _b(z) $ (see Eq.~\eqref{phi_b}). Two sensible choices would be placing the wall at $ z=0 $, the center of the bubble profile, and shifting it by its characteristic width to $ z=-L_w $.

\begin{figure}[h]
\centering
\includegraphics[width=0.75\linewidth]{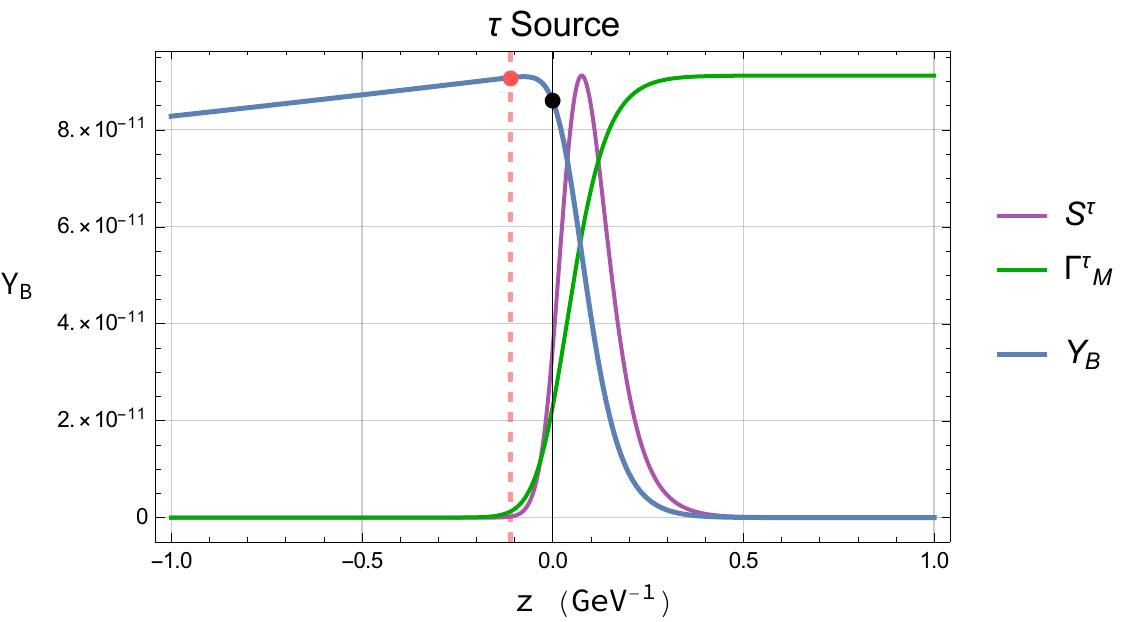}
\caption{$ Y_B $ with a tau source as a function of the position of the bubble wall. The blue line is $ Y_B $; the purple line is the source shape, arbitrarily normalized to fit plot scale; the green line is the relaxation rate, similarly normalized; and the dashed red line is the shift by the thickness of the bubble wall to $ z=-L_w = -0.11 \,\mathrm{GeV}^{-1} $. The black dot is the predicted $ Y_B $ placing the wall at $ z=0 $, and the red dot the predicted value $ Y_B = 9\times 10^{-11} $ placing the wall at $ z=-L_w $.
}
\label{fig:tauwlp}
\end{figure}

\begin{figure}[h]
\centering
\begin{subfigure}{.75\textwidth}
  \centering
  \includegraphics[width=1\linewidth]{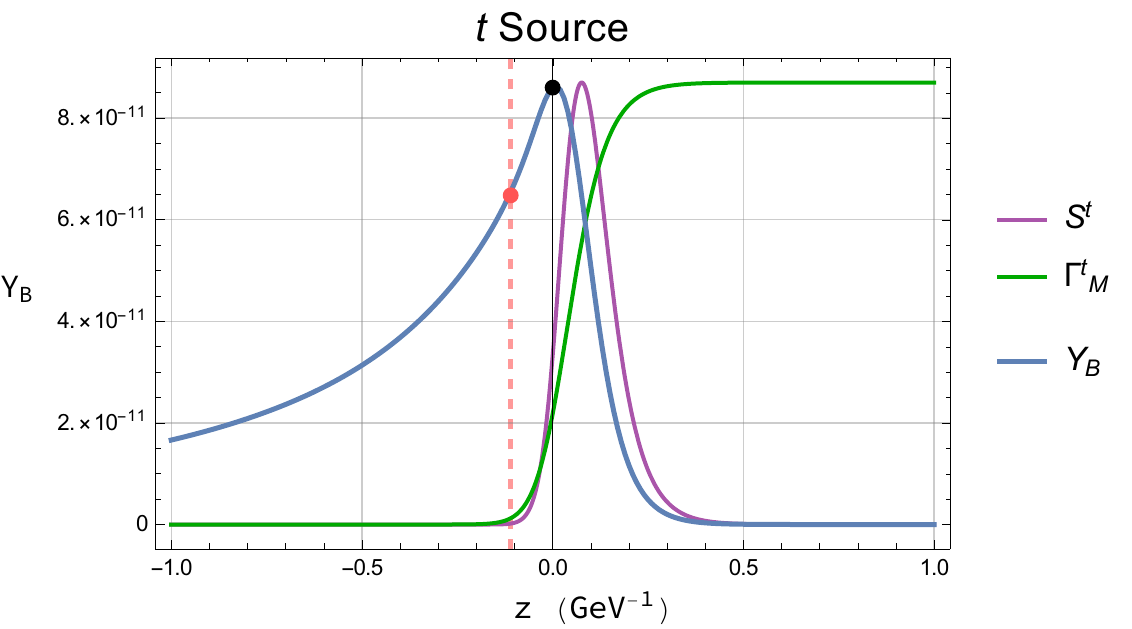}
\end{subfigure}\\
\begin{subfigure}{.75\textwidth}
  \centering
  \includegraphics[width=1\linewidth]{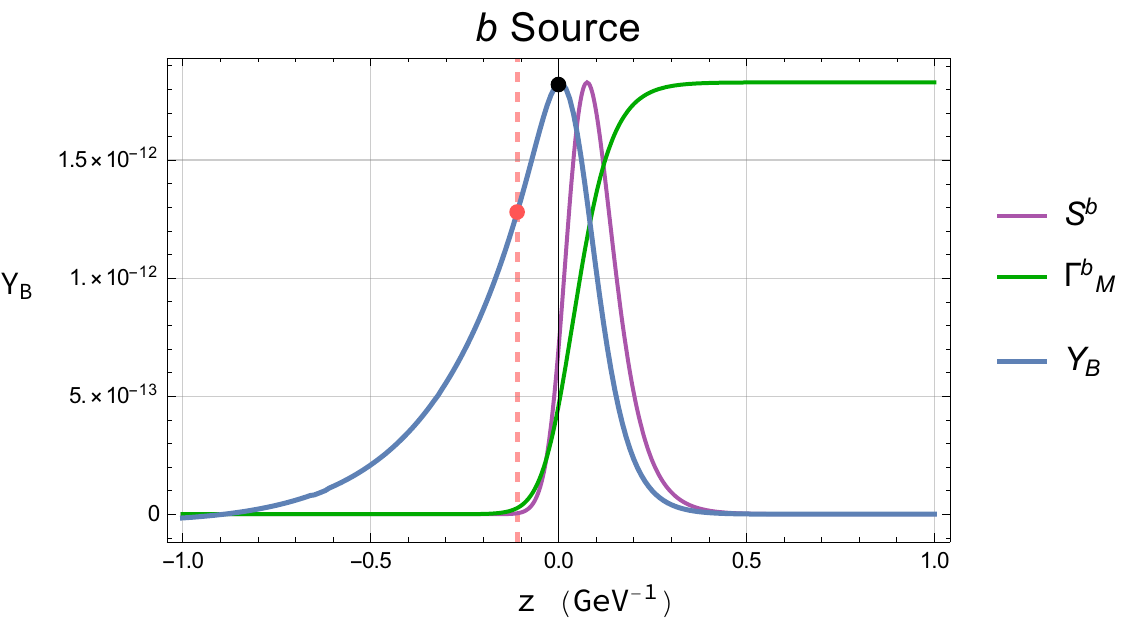}
\end{subfigure}
\caption{$ Y_B $ with a top source (upper plot) and a bottom source (lower plot) as a function of the position of the bubble wall, together with their sources $ S^{t/b} $ and $ \Gamma _M^{t/b} $, normalized to fit the plot.}
\label{fig:topwlp}
\end{figure}

In this section, we estimate the impact of this choice. In Figures~\ref{fig:tauwlp} and \ref{fig:topwlp}, we plot the baryon asymmetry obtained by shifting the point chosen for the step function. We overlay the plot of $ Y_B $ as a function of the wall shift with the shape of the source, which is maintained in this approximation, and with the shape of $ \Gamma _M $.

We can see that shifting the wall to the right quickly eliminates the generated baryon asymmetry. This is because the source is truncated: at a shift of $ +0.5 \,$GeV$^{-1}$, there is virtually no source left in the broken phase (recall that the source is taken with its $ z $ dependence, but taken as active only in the broken phase), and hence no baryon asymmetry. For negative shifts, the source is fully present in the broken phase, but we also overestimate the relaxation rates by taking the approximating step functions to be active in regions where the corresponding $ \Gamma _M $'s are in fact already highly suppressed. This explains the decrease in $ Y_B $ for negative shifts. The exact position of the peak is set by the competition between the inhibitory effect of overestimating $ \Gamma _M $ and the enhancement by including more of the source. We find that the variation in the predicted $ Y_B $ between placing the wall at $ z=0 $ and $ z=-L_w $ is $ \sim 5\% $ for $ \tau $,
$\sim 20\%$ for $t$ and $\sim 50\%$ for $b$.

\subsection{Bubble wall thickness and velocity}
\begin{figure}[h]
\centering
\includegraphics[width=0.7\linewidth]{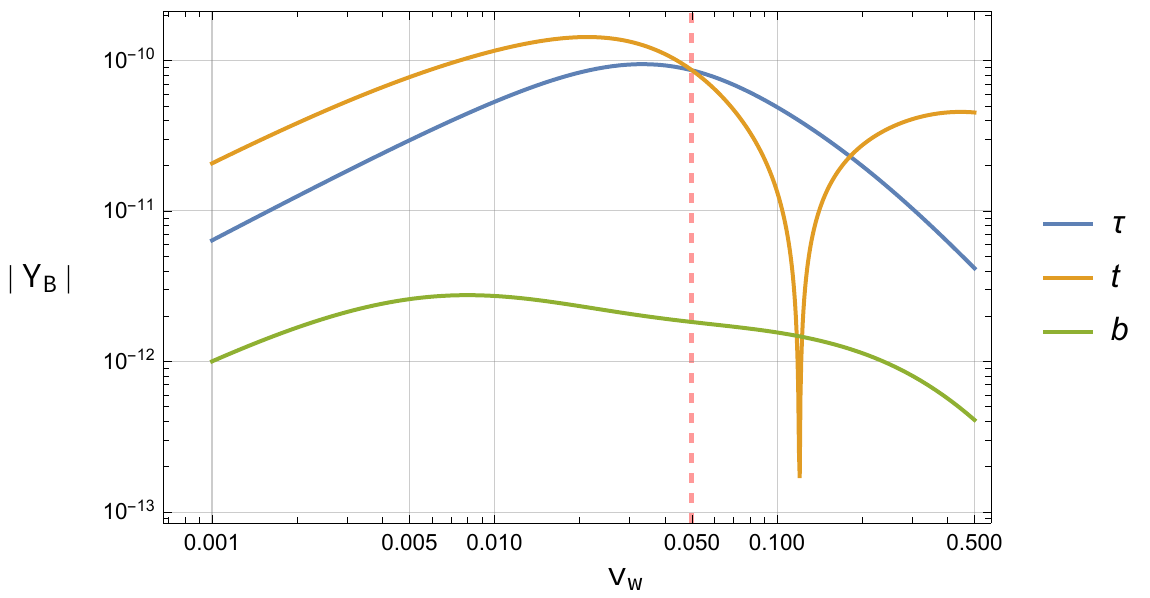}
\caption{The baryon asymmetry (in absolute value) $ \abs{Y_B} $ as a function of the bubble wall velocity $ v_w $. Each curve represents a source term. The benchmark value $ v_w = 0.05 $ is marked in dashed-red which is close to the optimal value of $ v_w $ for the top and the tau. The sharp dip in the curve for $ t $ (orange curve) is a point where $ Y_B $ changes sign. As explained in Table~\ref{tab:rates T=0} of Appendix~\ref{benchmarks}, the values for $ T_I^{\tau,t} $ are chosen such that $ Y_B = Y_B^{\mathrm{obs}} $ at $ v_w = 0.05 $, while $ T_I^b $ is arbitrarily normalized since it cannot produce the observed asymmetry.}
\label{fig:vwallp}
\end{figure}

\begin{figure}[h]
\centering
\includegraphics[width=0.7\linewidth]{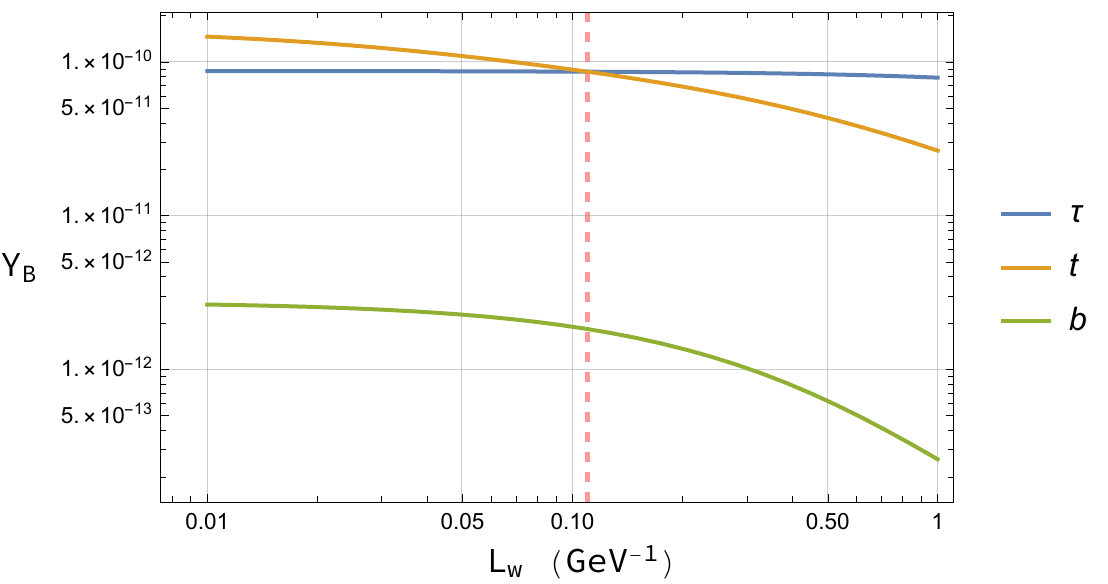}
\caption{The baryon asymmetry $ Y_B $ as a function of the bubble wall width $ L_w $. Each curve represents a source term. The benchmark value $ L_w = 0.11 \, \mathrm{GeV}^{-1}$ is marked in dashed-red. The choice of $ T_I^{\tau ,t,b} $ is the same as in Fig.~\ref{fig:vwallp}.
}
\label{fig:Lwallp}
\end{figure}

Successful EWBG requires a strong first order phase transition.  The details of the phase transition and the subsequent bubble nucleation and growth are important features that for each specific model will determine important parameters such as the wall thickness and wall velocity. Such a study is beyond the scope of the present paper, we refer the reader to recent analyses \cite{Kozaczuk:2015owa,Cline:2020jre}.
In our approach, we estimate the impact of modifying the bubble wall parameters: its wall velocity $ v_w $ and thickness $ L_w $. The wall velocity can directly impact the diffusion time scale for successful baryogenesis and the validity of the two-step approach. In Fig.~\ref{fig:vwallp}, we plot the baryon asymmetry as a function of the bubble wall velocity for each source, while in Fig.~\ref{fig:Lwallp}, we plot $Y_B^f$ as a function of the bubble wall width.
The numerical values of $T_I^f$ (see Tab.~\ref{tab:rates T=0}) are chosen such that $Y_B^t$ and $Y_B^\tau$ equal the observed baryon asymmetry for the benchmark values of $v_w$ and $L_w$ whereas $Y_B^b$ does not reach $Y_B^{\rm obs}$. \\
We see the importance of the parameters related to the phase transition in the large changes of the predicted asymmetry in response to changes in the wall velocity and width. The asymmetry from a tau source is less affected by $L_w $, varying only mildly from $Y_B^\tau(L_w = 0.01\,\mathrm{GeV}^{-1})\approx 8.7\times10^{-11}$ down to $Y_B^\tau(L_w=1\,\mathrm{GeV}^{-1}) \approx 7.8\times10^{-11} $, whereas the top- and bottom-sourced asymmetry depend more strongly on $L_w$, with a similar slope for $t$ and $b$. The change of sign in $Y_B^t(v_w)$ is yet another aspect of the sensitivity of the top source to model parameters.
While the benchmark value of $v_w=0.05$ is near-optimal for the $t$- and $\tau$-sources (cf. also Ref.~\cite{deVries:2018tgs}), the formalism of Ref.~\cite{Cline:2020jre} beyond the small-$v_w$ approximation shows that high yields of $Y_B$ are also possible for larger $v_w$. For large $ L_w $, the ultra-thin wall approximation (taking $ \Gamma _M^f $ as step functions) might also become less accurate, although important length scales as migration, diffusion and interaction lengths, as defined and discussed in Ref.~\cite{deVries:2018tgs}, are still larger than $ L_w^{\mathrm{max}} = 1 $ above.

\section{Conclusions and discussion}\label{sec: conclusions}
For the calculation of the baryon asymmetry of the Universe in electroweak baryogenesis, we developed a simple and useful method for solving the transport equations semi-analytically as well as fully analytically by exploiting various aspects of the structure of the set of these differential equations which couple the participating particle species. We obtained a physical picture of diverging and converging modes and identified the zero-modes as crucial components for the possibility of generating a non-zero baryon asymmetry. Maintaining the analytical form allowed us to identify important features and analytic dependence of the baryon asymmetry on model parameters.

While the derivation of our method is general, for the numerical evaluation we calculated the baryon asymmetry within the SMEFT framework with complex Yukawa couplings of the third-generation fermions and the muon.
We analyzed how modifications of model parameters and rates affect the resulting baryon asymmetry. This allowed us to estimate the sensitivity of the baryon asymmetry to changes by a factor of $ \mathcal O (10) $ that may result from more precise calculations of these parameters and rates. This large factor is chosen as a conservative example of modifications.

An important feature of our method is that it is straightforward to implement and avoids possible instabilities by the analytical reduction of the system before numerical evaluations are performed.
We confirmed the robustness of our method by the following consistency checks:
\begin{itemize}
\item Robustness to small changes in model parameters, such as the velocity and thickness of the bubble wall, as well as variations of the relaxation, Yukawa and sphaleron rates, with sensible dependence on the parameters. For reasonable values of the model parameters, we find no pathological behaviors. Furthermore, we investigated the impact of the ultra-thin wall approximation by varying the assumed position of the bubble wall.
\item Convergence of the one- and two-step approaches (that differ by the inclusion of the weak sphaleron rate in the transport equations) in the limit of a small weak sphaleron rate; with a relative difference of
$\sim 4,~15\%$ for the $\tau$, $t$, respectively, at the nominal weak sphaleron rate.
\item Good agreement between the semi-analytic and fully-analytic results in all the scenarios that can be tested with the fully-analytic method. The relative deviation remains below $\mathcal{O}\left(10^{-11}\right)$ for approximating all diffusion coefficients equal, and below $\mathcal{O}\left(10^{-4}\right)$ for distinguishing between a quark and a lepton diffusion constant.
\item Derivatives of particle densities receive the correct coefficients in the eigenvectors: precisely an extra factor of the eigenvalue, as expected by exponential solutions, up to relative differences of $\mathcal{O}\left(10^{-5}\right)$.
\item Summing over particle densities confirms $ B-L $ conservation
up to relative deviations of $\mathcal{O}\left(10^{-5}\right)$ or better.
\item Our method produces consistent results (within less than $1\%$) whether we incorporate or neglect light particles, as physically expected. This implies that it does not suffer from the increase in computational complexity when enlarging the $ K $ matrix. We checked this consistency by explicitly solving the transport equations for various set-ups of the full SM fermionic sector, which we used to produce the muon results in Ref.~\cite{Fuchs:2019ore}.
\end{itemize}

We conclude that the main conclusions presented in our previous works \cite{Fuchs:2020uoc,Fuchs:2019ore} are robust, even when considering the various approximations made and the large uncertainties in various parameters: A complex tau-Yukawa coupling can be the source of the CP violation that is required for electroweak baryogenesis, while complex top-, bottom-, and muon-Yukawa couplings can only account for a small part of it.

\acknowledgments{We are grateful to Jorinde van de Vis for very helpful discussions.
EF was supported by the Minerva Foundation.
ML would like to deeply thank the Weizmann Institute of Science for its hospitality during the completion of this work.
YN is the Amos de-Shalit chair of theoretical physics, and is supported by grants from the Israel Science Foundation (grant number 394/16), the United States-Israel Binational Science Foundation (BSF), Jerusalem, Israel (grant number 2014230), and the Yeda-Sela (YeS) Center for Basic Research.
}

\appendix

\section{Definitions and benchmark parameter values}\label{benchmarks}
In this Appendix we present the expressions and values for all parameters required to fully reproduce the final results.

\subsection{Benchmark parameters}
We take the nucleation temperature to be $ T_N = 88 $ GeV. At this temperature, the gauge couplings and Higgs VEV are given by \cite{deVries:2017ncy}
\begin{align}
g' &= 0.36 \,, & g &= 0.65 \,, & g_s &= 1.23 \,, & v_N &= 152 \, \text{GeV} \,.
\end{align}
The entropy density, written in terms of the temperature and the entropy degrees of freedom $ g^* $, is given by \cite{deVries:2017ncy}
\begin{align}
s = \frac{2\pi ^2}{45}g^* T_N^3 \,, \ \ \ g^* = 106.75 .
\end{align}
The bubble wall velocity and width are taken from \cite{deVries:2018tgs}, with values
\begin{align}
v_w = 0.05 \,, \ \ \ L_w = 0.11 \, \text{GeV}^{-1} \,.
\end{align}
The diffusion coefficients are approximately given by \cite{Joyce_1996,Cline:2000nw}
\begin{align}
D_{l_R} &= 380/T \,, & D_{l_L} &= 100/T \,, & D_u &= D_d = D_q = 6/T \,, & D_h = 100/T \,.
\end{align}

\subsection{Thermal properties}
The real part of the thermal mass of a particle $ f $ is of the form
\begin{align}
\text{Re}[\delta m_f^2(T)] = \left( \sum _i c_i g_i^2 + c_y y_f^2 \right)T^2 \,,
\end{align}
where $ g_i $ are the gauge couplings and $c_i$ are combinatorial coefficients. We denote a left (right) handed lepton by $ l_{L(R)} $, a left-handed quark doublet by $ q $ and a right-handed up (down) type quark by $ u \ (d) $. The thermal masses are given by \cite{Enqvist_1998}
\begin{align}\label{thermal masses}
\begin{split}
\text{Re}[\delta m_{l_R}^2(T)] &= \left( \frac{1}{8}g_y^2 + \frac{1}{8}y_{l_R}^2 \right)T^2 \,,\\
\text{Re}[\delta m_{l_L}^2(T)] &= \left( \frac{3}{32}g_w^2 + \frac{1}{32}g_y^2 + \frac{1}{16}y_{l_R}^2 \right)T^2 \,,\\
\text{Re}[\delta m_u^2(T)] &= \left( \frac{1}{6}g_s^2 + \frac{1}{18}g_y^2 +\frac{1}{8}y_u^2 \right)T^2 \,,\\
\text{Re}[\delta m_d^2(T)] &= \left( \frac{1}{6}g_s^2 + \frac{1}{72}g_y^2 +\frac{1}{8}y_d^2 \right)T^2 \,,\\
\text{Re}[\delta m_q^2(T)] &= \left( \frac{1}{6}g_s^2 + \frac{3}{32}g_w^2 + \frac{1}{288}g_y^2 +\frac{1}{16}y_u^2 +\frac{1}{16}y_d^2 \right)T^2 \,,\\
\text{Re}[\delta m_h^2(T)] &= \left( \frac{3}{16}g_w^2 + \frac{1}{16}g_y^2 +\sum _{i,j} \left( \frac{1}{12}y_{l_R^i}^2 +\frac{1}{4}y_{u^j}^2 +\frac{1}{4}y_{d^j}^2 \right) \right)T^2 \,.
\end{split}
\end{align}
The $ k $-functions related to the chemical potentials in Eq. \eqref{chemical potentials} are calculated as \cite{deVries:2017ncy}
\begin{align}
k_f(a_f) &= \tilde k^f\frac{c_{F/B}}{\pi ^2}\int _{a_f}^\infty \dx \frac{xe^x}{(e^x\pm 1)^2}\sqrt{x^2 - a_f^2} \ , & \left( a_f \equiv \sqrt{\text{Re}[\delta m_f^2(T)]}/T \right)
\end{align}
where $\tilde k^f$ counts the physical degrees of freedom in the multiplet (e.g. $ \tilde k^q = 6 $, $ \tilde k^H = 4 $), $ c_{F(B)} = 6(3) $, and $ +(-) $ is chosen for fermions (bosons).

The thermal widths are given by \cite{Elmfors_1999}
\begin{align}
\Gamma _{\text{lepton}} \approx 0.002 T \,, \ \ \ \Gamma _{\text{quark.}} \approx 0.16 T \,.
\end{align}
Next, we define
\begin{align}
\omega _{R/L}^f(\textbf{k}) = \sqrt{\abs{\textbf{k}}^2 + \text{Re}[\delta m_{f_{R/L}}^2(T)]} \,, \ \ \ \mathcal E _{R/L}^f(\textbf{k}) = \omega _{R/L}^f(\textbf{k}) - i\Gamma_f \,,
\end{align}
and
\begin{align}
n_F(k_0) = \frac{1}{e^{k_0/T}+1} \,, \ \ \ h(k_0) = \frac{e^{k_0/T}}{(e^{k_0/T}+1)^2} \,.
\end{align}
These are used to calculate the CPV source and the \CP-conserving rates. We use the kink solution as a typical ansatz for the space-dependent Higgs VEV:
\begin{align}\label{phi_b}
\phi _b(z) = \frac{v_N}{2}\left( 1 + \tanh \frac{z}{L_w} \right) .
\end{align}

\subsection{Source and \CP-conserving rates}\label{app:rates}
The \CP-violating source is proportional to the relative phase between the mass and its spatial derivative. Explicitly, the source is given by the expression \cite{Lee:2004we,Cirigliano_2006}
\begin{align}
S_f(z;T) &= \frac{v_wN_c^f}{\pi ^2}\text{Im}(m_f'm_f^* )J_f(T) = \frac{v_w N_c^f {Y_{SM}^f}^{\!\!\!\!2}}{\pi ^2 v_0^2} \frac{T_I^f}{\left(1+T_R^f\right)^2 + {T_I^f}^2}J_f(T) \phi _b^3(z)\phi _b'(z) ,\\
J_f(T) &= \int _0^\infty \frac{k^2dk}{\omega _L\omega _R}\text{Im} \left[ \frac{n_F(\mathcal E_L) - n_F(\mathcal E_R^*)}{(\mathcal E_L - \mathcal E_R^*)^2}(\mathcal E_L\mathcal E_R^* - k^2) + \frac{n_F(\mathcal E_L) + n_F(\mathcal E_R)}{(\mathcal E_L + \mathcal E_R)^2}(\mathcal E_L\mathcal E_R + k^2) \right] .\nonumber
\end{align}

For the relaxation and Yukawa rates of the \CP-conserving processes, we neglect hole modes to get
\begin{align}\label{Gamma M,Y expression}
\begin{split}
\Gamma _M^f &= \frac{3N_c^f}{\pi ^2T^3} \abs{m_N^f}^2 \!\int _0^\infty \frac{k^2dk}{\omega _L\omega _R}\text{Im} \left[ \frac{h(\mathcal E_L) + h(\mathcal E_R)}{\mathcal E_R + \mathcal E_L}(\mathcal E_L\mathcal E_R + k^2) - \frac{h(\mathcal E_L) + h(\mathcal E_R^*)}{\mathcal E_R^* - \mathcal E_L}(\mathcal E_L\mathcal E_R^* - k^2)\right],\\
\Gamma _Y^f &= \Gamma _Y^{f,3} + \Gamma _Y^{f,4},
\end{split}
\end{align}
where 
\begin{align}
\begin{split}
\Gamma _Y^{f,3} &= \frac{3N_cY_f^2}{4\pi ^3T^2}(m_L^2 + m_R^2 - m_H^2)\int _{m_R}^\infty d\omega _R h(\omega _R)\times\\
&\quad \ \bigg[\ln \left[ \frac{e^{-\beta \omega _R}+e^{\beta \omega _-}}{e^{-\beta \omega _R}+e^{\beta \omega _+}}\frac{e^{\beta \omega _+}-1}{e^{\beta \omega _-}-1} \right]\theta (m_L - m_R - m_H)\\
&\ \ \, + \ln \left[ \frac{e^{\beta \omega _R}+e^{\beta \omega _-}}{e^{\beta \omega _R}+e^{\beta \omega _+}}\frac{e^{\beta \omega _+}-1}{e^{\beta \omega _-}-1} \right]\left( \theta (m_R - m_L - m_H) - \theta (m_H - m_L - m_R) \right)\bigg]\\
\Gamma _Y^{f,4} &= \frac{\zeta_3}{6\pi ^3}g_s^2Y_f^2 T \ln \left[ \frac{8T^2}{\text{Re}[\delta m_f^2(T)]} \right],\\
\omega _\pm &= \frac{1}{2m_R^2}\bigg[\omega _R \abs{m_H^2  + m_R^2 - m_L^2}\\
&\quad \, \pm \sqrt{(\omega _R^2 - m_R^2)(m_R^2 - (m_L+m_H)^2)(m_R^2 - (m_L-m_H)^2)}\bigg]\,.
\end{split}
\end{align}
$m_N$ is the mass at the nucleation temperature, determined by the kink solution, $ \zeta_3 \approx 1.202 $, $ N_c $ is the number of colors, and $ m_R,m_L, $ and $ m_H $ are short for the thermal masses \eqref{thermal masses}. The leading contribution to $\Gamma_Y^f$ contains an external gluon line. Although it is not strictly a Yukawa interaction, gauge fields are taken to be in equilibrium and are not part of the transport equations.
We approximate the rates as independent constants in each phase. In the symmetric phase, we consider all $ \Gamma_M^S[f] $ to vanish, while the Yukawa rates are approximately equal in both phases.

In Table~\ref{tab:rates T=0} we present the numerical values for $ \Gamma _M^B$ (in the broken phase) and $\Gamma ^Y $ when $ T_R = T_I = 0 $. We also present the benchmark values of $ T_I $ used throughout the text ($ T_R^{\mathrm{bench}} = 0 $ for all species). The values for $ t,\tau $ are chosen to reproduce $ Y_B = Y_B^{\mathrm{obs}} $. Since $ b,\mu $ cannot produce the observed asymmetry within collider bounds as single sources \cite{Fuchs:2019ore,Fuchs:2020uoc}, $ T_I^{\mathrm{bench}} $ for $ b,\mu $ are set to $ -0.05 $.
\setlength{\tabcolsep}{10pt}
\renewcommand{\arraystretch}{0.9}
\begin{table}[h]
\centering
\begin{tabular}{c|l l l}
\hline \hline
Particle & $ \Gamma_M^B \, (\text{GeV}) $ & $ \Gamma_Y \, (\text{GeV})$ & $ T_ I^{\mathrm{bench}} $\\ \hline
$ \tau $ & $ 4.9 \times 10^{-3} $ & $ 5.6 \times 10^{-4} $ &  $ -0.04363 $\\
$ \mu  $ & $ 1.7 \times 10^{-5} $ & $ 2.0 \times 10^{-6} $ & $ -0.05 $ \\
$ t $ & 102 & 2.6 & 0.019455 \\
$ b $ & $ 5.3 \times 10^{-2} $ & $ 1.7 \times 10^{-3} $ & $ -0.05 $ \\
\hline \hline
\end{tabular}
\caption{Numerical values for $ \Gamma _M^B$ (in the broken phase), $\Gamma _Y$ (for the both phases) as given by Eq. \eqref{Gamma M,Y expression}, with $ T_R=T_I=0 $, and the benchmark value for $ T_ I^{\mathrm{bench}} $.}
\label{tab:rates T=0}
\end{table}\\
For $ T_{R,I} \neq 0 $, the expressions in Eq.~\eqref{Gamma M,Y expression}, and hence the values in Table~\ref{tab:rates T=0}, get corrected according to
\begin{align}\label{gamma rescale}
\Gamma_M &\rightarrow \left[\frac{(1+r_{N0}^2T_R^f)^2+r_{N0}^2T_I^{f2}}{(1+T_R^f)^2+T_I^{f2}}\right] \Gamma_M\,,\nonumber\\
\Gamma_Y &\rightarrow \left[\frac{(1+3 r_{N0}^2T_R^f)^2+ (3r_{N0}^2T_I^f)^2}{(1+T_R^f)^2+T_I^{f2}}\right] \Gamma_Y\,.
\end{align}
Here $ r_{N0} \equiv v(T=T_N)/v(T=0) $, where $T_N$ is the nucleation temperature.

The sphaleron rates are estimated via lattice calculations, and are given by \cite{Bodeker:1999gx,Moore:2010jd}
\begin{align}
\Gamma _{\text{ws}} = 120 \alpha _w^5 T \approx 4.5\times 10^{-4} \,\text{GeV} \,, \ \ \ \Gamma _{\text{ss}} = 14 \alpha _s^4 T \approx 0.26 \,\text{GeV}\,.
\end{align}

\section{Consistency checks}\label{app:cons check}
\subsection{\textit{B--L} conservation}\label{app:B-L}
A simple and important check using the one-step approach is to verify that $B-L$ is conserved. We define the relative difference between the baryon and lepton numbers, as
\begin{align}
R_{B-L}= 2\abs{\frac{n_B-n_L}{n_B + n_L}} .
\end{align}
In Table~\ref{tab:rel diff} we show $ R_{B-L} $ for each of the four fermions of interest, setting $ T_R^f = 0, \ T_I^f = \pm 0.05$ (+ for $t$, - for $b, \tau, \mu$) in each case, and the rest of the dim-6 operators to zero. We find that across the parameter space the relative difference does not exceed $\sim 10^{-5}$.

\subsection{Derivative test}\label{app:derivative}
We construct our solution as a set of 1st order differential equations. Thus half of the entries are the first derivatives of the various particle densities. Recalling that the solutions are exponents, the entries of the derivative terms in each eigenvector should be the same as those of the corresponding particles, multiplied by the appropriate eigenvalue. In the fully analytic case, the equality is exact. In the semi-analytic case, we define the relative difference between a derivative entry and the particle entry times the appropriate eigenvalue as
\begin{align}
R_{f',f} = \max_{(f,i)} 2\abs{\frac{f_i' - \lambda_if_i}{f_i' + \lambda_if_i}}.
\end{align}
Here $ f_i $ denotes the entry of the $ i $'th eigenvector corresponding to particle $ f $. In Table~\ref{tab:rel diff} we show $ R_{f',f} $ for each dim-6 operator, in the broken phase. In each case, $ R_{f,f'} $ denotes the largest value of all particles and all eigenvectors.

\subsection{Number of particles}\label{app:nparticles}
Increasing the number of participating particles in the transport equations may result in numerical instabilities. Physically, very light particles should not affect the resulting prediction for the baryon asymmetry, and are typically neglected. Light quarks participate in strong sphaleron interactions, which are efficient at high temperatures. However, in the approximation that first and second generation quarks are massless and weakly interacting, they are degenerate in the transport equations, and we may choose a single representative to capture their contribution (see Sect.~\ref{sec: con and gen sol} for the explicit treatment; see also Ref.~\cite{deVries:2018tgs}). We verified that our method is robust to changing the number of participating fermions. We quote in Tab.~\ref{tab:rel diff} the resulting $ Y_B $ in two scenarios:
\begin{enumerate}
\item A set containing $ t,b,\tau ,u,h $, where $ u $ is a representative of the light quarks, which we used to produce the results in \cite{Fuchs:2020uoc}. Note that the muon does not appear here. We denote this scenario as $ Y_B^{t,b,\tau } $.
\item The full SM set, as used to produce the results for the muon \cite{Fuchs:2019ore}. We denote this as $ Y_B^{\text{all}} $.
\end{enumerate}
We set $ T_R^f = 0 , T_I^f = \pm 0.05 $ ($ + $ for $ t $, $ - $ for $ b,\tau,\mu $), and the rest of the dim-6 operators to zero.
\setlength{\tabcolsep}{10pt}
\renewcommand{\arraystretch}{0.9}
\begin{table}[h]
\begin{center}
\begin{tabular}{c|llll}
\hline \hline
Particle & $ t $ & $ b $ & $ \tau $ & $ \mu $ \\\hline
$ R_{B-L} $ & $ 1.0\times 10^{-5} $ & $ 1.1\times 10^{-6} $ & $ 1.5\times 10^{-8} $ & $ 5.3\times 10^{-8} $ \\
$ R_{f',f} $ & $ 6.4\times 10^{-7} $ & $ 1.4\times 10^{-6} $ & $ 9.8\times 10^{-7} $ & $ 9.2\times 10^{-6} $ \\ \hline
$ Y_B^{t,b,\tau } $ & $ 2.21\times 10^{-10} $ & $ 1.824\times 10^{-12} $ & $ 9.852\times 10^{-11} $ & $ - $ \\
$ Y_B^{\text{all}} $ & $ 2.20\times 10^{-10} $ & $ 1.820\times 10^{-12} $ & $ 9.846\times 10^{-11} $ & $ 1.0\times 10^{-12} $ \\
\hline \hline
\end{tabular}
\caption{For each fermion:
Relative difference between baryon and lepton number (first row) and between an eigenvector of a particle density and its derivative (second row);
prediction of $Y_B$ with $t,b,\tau,u,h$ (third row) participating in the transport equations, and with all SM particles including $\mu$ (fourth row), for $ T_R^f = 0 , T_I^f = \pm 0.05 $.}
\label{tab:rel diff}
\end{center}
\end{table}

\subsection{Comparing the one step and two step approaches}
\label{app:12step}
\begin{figure}[h]
\begin{subfigure}{.49\textwidth}
\centering
\includegraphics[width=1\linewidth]{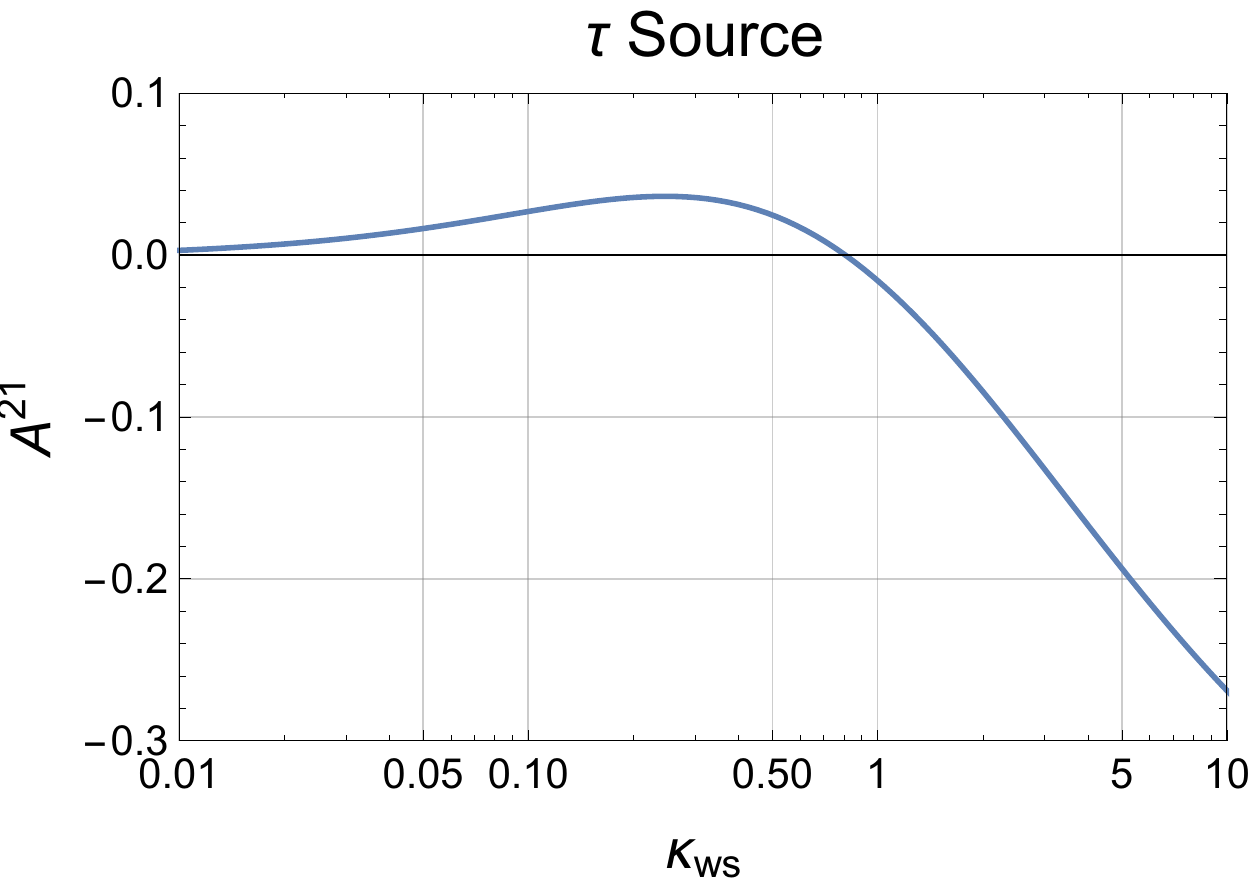}
\end{subfigure}\ \
\begin{subfigure}{.49\textwidth}
\centering
\includegraphics[width=1\linewidth]{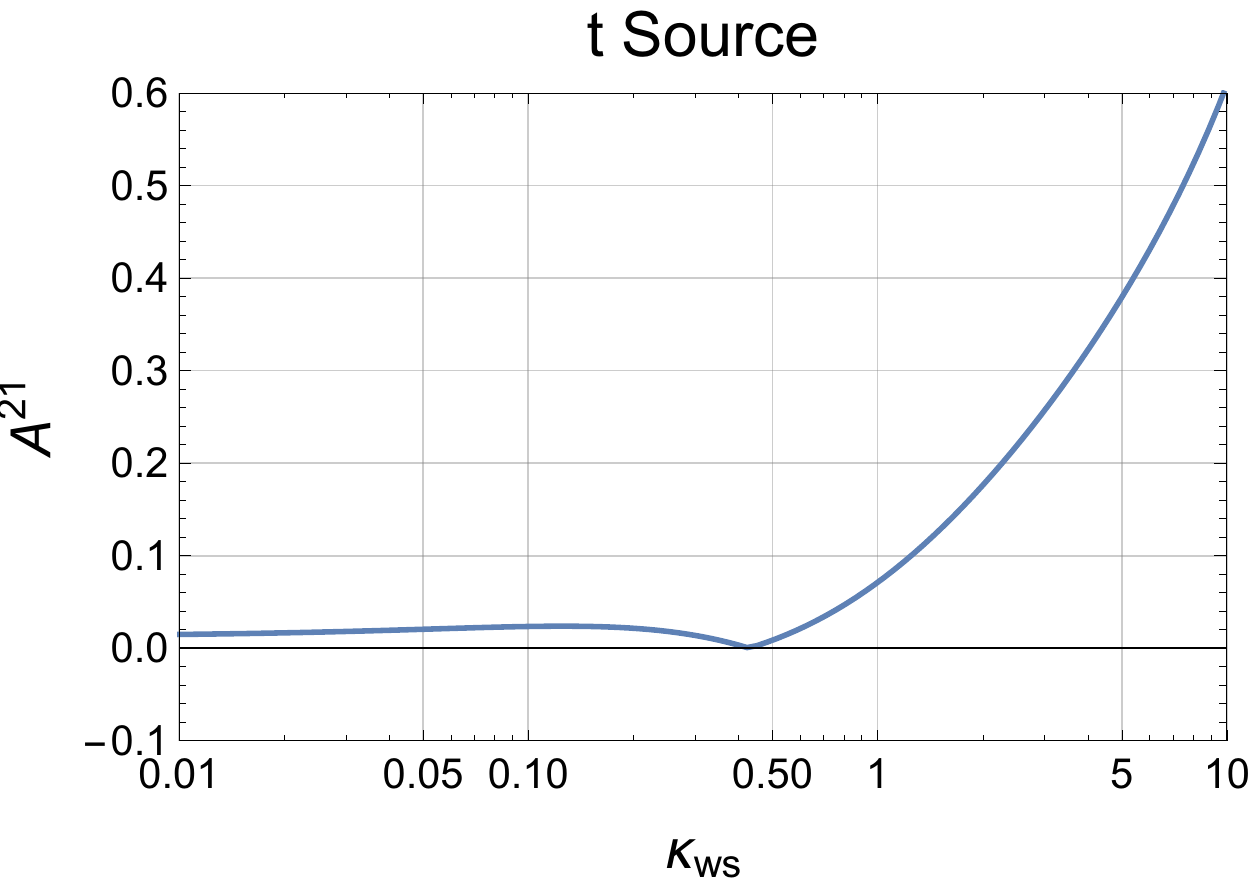}
\end{subfigure}
\caption{The relative difference $ A^{21} $ between the one- and two-step approaches as a function of the weak sphaleron rate. As $ \Gamma _{\text{ws}} $ is lowered, $ A^{21} $ vanishes.}
\label{fig:gws}
\end{figure}

Varying the weak sphaleron rate can be used to compare the one- and two-step approaches.
For the full SM set of transport equations,
we parameterize the difference in their predicted baryon asymmetry as a function of the weak sphaleron modifier $ \kappa_{\text{ws}} $, by
\begin{align}
A^{21}(\kappa_{\text{ws}}) = \frac{Y_B^2 - Y_B^1}{Y_B^2 + Y_B^1} \,,
\end{align}
where $ Y_B^1,Y_B^2 $ are the predicted baryon asymmetries as obtained in the one- and two-step approaches, respectively.
Since the one- and two step solutions differ by the inclusion of the weak sphaleron rate in the transport equations, we expect the two approaches to converge as we decrease the rate of the weak sphaleron, i.e. $ A^{21} \xrightarrow{\Gamma _{\text{ws}}\to 0} 0 $, and indeed we see this behavior in Fig.~\ref{fig:gws}.
\\
For the benchmark value $ \Gamma _{\text{ws}} $ ($ \kappa _{\text{ws}} = 1 $) we obtain
\begin{align} \label{eq:A21kappaws1}
A^{21}_\tau (1) \sim 0.02 ,\ \ \ A^{21}_t (1) \sim 0.07 ,
\end{align}
which corresponds to a deviation of $ \sim 4,~15\% $, respectively. At $ \kappa _{\text{ws}} = 0.01 $, we find
\begin{align}
A^{21}_\tau(0.01) \sim 0.003, \ \ \
A^{21}_t(0.01) \sim 0.01.
\end{align}
The relative difference is small at the literature value corresponding to $\kappa_{\text{ws}}=1$. It grows for large values of $ \Gamma _{\text{ws}}$, but remains below $30\%$ for the $t$ and $60\%$ for the $\tau$ at $\kappa_{\text{ws}}=10$. Hence the two-step approach still reproduces the order of magnitude of $Y_B^1$ even in the extreme case of such a large correction factor of the weak sphaleron rate.
\subsection{Comparing the semi-analytic and fully analytic methods}\label{app:semi-analytic}
Under certain approximations, we can solve the transport equations analytically. Below is a comparison of the semi analytic method to the fully analytic method in scenarios where it is applicable. We consider the two scenarios described in Section~\ref{sec: block decomp}:
\begin{itemize}
\item \textit{Case 1}: We neglect the Higgs density, and approximate the diffusion coefficients as equal among all fermions, left and right, setting $ D_f = 100/T $.
\item \textit{Case 2}: We neglect the Higgs density, but the diffusion coefficients are taken to be equal separately among quarks and leptons, i.e.
\begin{align*}
\begin{cases}
D_{q_L} = D_{q_R} = 6/T \,,\\
D_{l_L} = D_{l_R} = 100/T \,.
\end{cases}
\end{align*}
\end{itemize}
In Table~\ref{tab:rel diff 2} we show the largest deviations in the eigenvalues $ \lambda _i $\, between the semi-analytic and the fully analytic solution, eigenvector entries $ \phi _{ij} $ and baryon asymmetry $ Y_B $. All these quantities are calculated for a tau source\footnote{Changing the active dim-6 operator has a very mild impact on the eigensystem. We quote $ R $'s for the eigenvalues and eigenvectors in the broken phase.}. We define $ R_\lambda \,,\ R_\phi \,,\ R_{Y_B^\tau } $\,, respectively, similarly to the definition of the $ B-L $ conservation and derivative test.
\setlength{\tabcolsep}{12pt}
\renewcommand{\arraystretch}{0.9}
\begin{table}[h]
\begin{center}
\begin{tabular}{l|lll}
\hline\hline
Test & $ R_\lambda $ & $ R_\phi $ & $ R_{Y_B^\tau } $ \\
\hline
Case 1: $D_f$ & $ 2.0\times 10^{-13} $ & $ 5.9\times 10^{-14} $ & $ 5.8\times 10^{-12} $ \\
Case 2: $D_q, D_l$ & $ 1.8\times 10^{-12} $ & $ 4.5\times 10^{-9} $ & $ 1.1\times 10^{-4} $ \\
\hline\hline
\end{tabular}
\caption{Largest relative differences between the semi- and fully-analytic approaches for the eigenvalues $\lambda$, eigenvectors $\phi$, and the baryon asymmetry; approximating a universal diffusion constant for all fermions (case 1), or one diffusion constant for all quarks and one for all leptons (case 2).}
\label{tab:rel diff 2}
\end{center}
\end{table}\\
We see that the error remains small throughout the calculation, and the resulting baryon asymmetry is in good agreement between the semi- and fully-analytic procedures. We also checked the case $ D_l = 380/T $
and $ D_\tau = 100/T $, following the direct calculation outlined in Section~\ref{sec: block decomp}. The results are very similar to Case 2, thus we do not show them here explicitly.

\section{The baryon asymmetry}\label{app: BAU}
For completeness, we present the derivation of the expression for the baryon asymmetry, assuming the chiral density $n_L$ has already been solved for. The solution for the transport equations is then a straightforward generalization to this procedure.

We approximate the dynamics of the baryon density, $n_b$, by a one-dimensional differential equation in the bubble wall frame, placing a planar wall at $ z=0 $, with the broken phase chosen to be $ z>0 $.
Using the diffusion approximation, similarly to the transport equations, the equation for the baryon density $ n_b $ is
\begin{align}\label{nb ode}
n_b''(z) - \frac{v_w}{D_q}n_b'(z)= \frac{\Gamma _{ws}(z)}{D_q}\left(\mathcal R n_b(z) + \frac{3}{2}n_L(z) \right) \equiv \frac{\Gamma _{ws}(z)}{D_q}\mathcal R n_b + f(z),
\end{align}
where $ v_w $ is the bubble wall velocity, $ D_q $ is the quark diffusion coefficient, $ \Gamma _{\text{ws}} $ is the weak sphaleron rate and $ \mathcal R = 15/4 $ is the so-called SM relaxation term.
The sphaleron process is efficient only in the symmetric phase \cite{Bodeker:1999gx,Moore:2010jd,DOnofrio:2014rug}(assuming a strongly first order phase transition), and we therefore take the sphaleron rate to be a step function $ \Gamma _{\text{ws}} \to \Gamma _{\text{ws}}\theta (-z) $, where $ \Gamma _{\text{ws}} $ is constant. All other coefficients are constant numbers as well,  and the chiral density acts as an external source for the baryon number density, which, due to the sphaleron rate, is active only in the symmetric phase.

In the broken phase, the solution to Eq. \eqref{nb ode} is of the simple form
\begin{align}\label{nb sol broken}
n_b(z) = A_1 + A_2e^{v_wz/D_q} \,,
\end{align}
while in the symmetric phase the homogeneous solution is of the form
\begin{align}\label{nb sol sym hom}
\begin{split}
n_b^h(z) &= B_1e^{\frac{z}{2D_q}\left( v_w - \sqrt{4D_q\Gamma _{\text{ws}} \mathcal R + v_w^2} \right)} + B_2e^{\frac{z}{2D_q}\left( v_w + \sqrt{4D_q\Gamma _{\text{ws}} \mathcal R + v_w^2} \right)}\\
&\equiv B_1e^{\alpha _-z} + B_2e^{\alpha _+z} \equiv B_1u_1(z) + B_2u_2(z)\,.
\end{split}
\end{align}
The particular solution is obtained by variation of parameters
\begin{align}
n_b^p(z) = K_1(z)u_1(z) + K_2(z)u_2(z) \,.
\end{align}
Using the Wronskian
\begin{align}
\begin{split}
W &= \begin{vmatrix}
e^{\alpha _-z} & e^{\alpha _+z}\\
\alpha _-e^{\alpha _-z} & \alpha _+e^{\alpha _+z}
\end{vmatrix}= (\alpha _+ - \alpha _-)e^{(\alpha _+ + \alpha _-)z} \\
&= \frac{1}{D_q}\sqrt{4D_q\Gamma _{\text{ws}}\mathcal R + v_w^2}e^{(\alpha _+ + \alpha _-)z}\equiv \frac{k}{D_q}e^{(\alpha _+ + \alpha _-)z} \,,
\end{split}
\end{align}
we solve
\begin{align}
\begin{split}
K_1(z) &= -\int \dz\frac{1}{W(z)}u_2(z)f(z) = - \frac{D_q}{k}\int e^{-(\alpha _- + \alpha _+)z}e^{\alpha _+z}f(z)\dz\\
&= -\frac{3\Gamma _{ws}}{2k}\int_0^z \theta (-x)e^{-\alpha _-x}n_L(x)\dx \,,\\
K_2(z) &= \int \dz\frac{1}{W(z)}u_1(z)f(z) = \frac{3\Gamma _{ws}}{2k}\int_0^z \theta (-x)e^{-\alpha _+x}n_L(x)\dx \,.
\end{split}
\end{align}
The particular solution is thus given by
\begin{align}\label{nb sol sym p}
n_b^p(z) &= \frac{3\Gamma _{ws}}{2k}\left[ e^{\alpha _+z}\int_0^z \theta (-x)e^{-\alpha _+x}n_L(x)\dx - e^{\alpha _-z}\int_0^z \theta (-x)e^{-\alpha _-x}n_L(x)\dx \right].
\end{align}

Let us impose boundary conditions. In the broken phase, the second term in Eq. \eqref{nb sol broken} diverges as $ z\to \infty $, and we set $ A_2 = 0 $. The baryon number density is thus completely determined by $ A_1 $ and therefore by the continuity condition at $ z=0 $. In the symmetric phase, the second term in Eq. \eqref{nb sol sym hom} vanishes as $ z\to -\infty $, but the first one diverges. In  Eq. \eqref{nb sol sym p}, the first term vanishes while the second terms diverges.
The divergence of the second term is manifest,
and in App.~\ref{app:proof} we show that the first one does indeed vanish.
For the divergent term, we set
\begin{align}\label{nb B1 = integral}
B_1 = \frac{3\Gamma _{ws}}{2k}\int_0^{-\infty} e^{-\alpha _-x}n_L(x)\dx \,,
\end{align}
such that
\begin{align}\label{nb divergent mode cancellation}
\lim_{z\rightarrow -\infty } \left( B_1 - \frac{3\Gamma _{ws}}{2k}\int_0^z e^{-\alpha _-x}n_L(x)\dx \right)e^{\alpha _-z} = \lim_{z\rightarrow -\infty } \frac{3\Gamma _{ws}}{2k}\int_z^{-\infty} e^{\alpha _-(z-x)}n_L(x)\dx = 0 \,.
\end{align}
Consider now the derivative of Eq. \eqref{nb sol sym p}
\begin{align*}
{n_b^p}'(z) = \frac{3\Gamma _{ws}}{2k}\bigg[ \alpha _+e^{\alpha _+z}\int_0^z \theta (-x)e^{-\alpha _+x}n_L(x)\dx - \alpha _-e^{\alpha _-z}\int_0^z \theta (-x)e^{-\alpha _-x}n_L(x)\dx\\
+e^{\alpha _+z} \theta (-z)e^{-\alpha _+z}n_L(z) - e^{\alpha _-z} \theta (-z)e^{-\alpha _-z}n_L(z) \bigg] \,.
\end{align*}
The second line is identically zero, and the first vanishes at $ z=0 $. From this we obtain the continuity condition for $ n_b,n_b' $ at $ z=0 $:
\begin{align*}
A_1 = B_1 + B_2 \,, \ \ \ B_2 = -\frac{\alpha _-}{\alpha _+}B_1 \,,
\end{align*}
and finally we obtain
\begin{align}\label{Y_B}
\begin{split}
Y_B &= \frac{n_b(z>0)}{s} = \frac{A_1}{s} = \frac{1}{s}\left( 1-\frac{\alpha _-}{\alpha _+} \right)B_1 = \frac{k}{D_q\alpha _+s}B_1\\
&= \frac{3\Gamma _{ws}}{2D_q\alpha _+s}\int_0^{-\infty} e^{-\alpha _-x}n_L(x)\dx \,.
\end{split}
\end{align}

\subsection{Vanishing term in particular solution} \label{app:proof}
Here we prove that the first term of Eq.~(\ref{nb sol sym p}) in the derivation of the particular solution of the baryon density, $n_b^p(z)$, in Sect.~\ref{app: BAU} vanishes under the boundary conditions.
For simplicity we flip $z\to-z$ and consider instead the limit $z\to\infty$.
Let $\lambda$ be a positive constant, and suppose $ f(x) $ is positive and bounded, and that $ \int _0^\infty f(x)\dx $ converges. We wish to show that
\begin{align*}
\lim_{z\to  \infty }e^{-\lambda z}\int _0^z e^{\lambda x}f(x)\dx = 0 \,.
\end{align*}
Choose some constant $ a<z $, then
\begin{align}
\begin{split}
\int _0^z e^{\lambda (x-z)}f(x)\dx &= \int _0^{z-a} e^{\lambda (x-z)}f(x)\dx + \int _{z-a}^z e^{\lambda (x-z)}f(x)\dx \\
&\leqslant \int _0^{z-a} e^{\lambda (x-z)}f(x)\dx + \int _{z-a}^z f(x)\dx\\
&\leqslant e^{-\lambda a}\int _0^\infty f(x)\dx + \int _{z-a}^\infty f(x)\dx \,.
\end{split}
\end{align}
We eliminated the exponent in the second term because $ e^{\lambda (x-z)} \leqslant 1 $, and in the first term we have $ e^{\lambda (x-z)} \leqslant e^{\lambda ((z-a)-z)} = e^{-\lambda a} $. The second term clearly vanishes as $ z\to \infty $ because $ \int _0^\infty f(x)\dx $ is finite. The first term is a constant times $ e^{-\lambda a} $. Since $ a $ is arbitrary, the first term is an arbitrarily small upper bound, and the limit is 0.

\bibliographystyle{apsrev4-1}
\bibliography{References}

\begin{thebibliography}{33}%
\makeatletter
\providecommand \@ifxundefined [1]{%
 \@ifx{#1\undefined}
}%
\providecommand \@ifnum [1]{%
 \ifnum #1\expandafter \@firstoftwo
 \else \expandafter \@secondoftwo
 \fi
}%
\providecommand \@ifx [1]{%
 \ifx #1\expandafter \@firstoftwo
 \else \expandafter \@secondoftwo
 \fi
}%
\providecommand \natexlab [1]{#1}%
\providecommand \enquote  [1]{``#1''}%
\providecommand \bibnamefont  [1]{#1}%
\providecommand \bibfnamefont [1]{#1}%
\providecommand \citenamefont [1]{#1}%
\providecommand \href@noop [0]{\@secondoftwo}%
\providecommand \href [0]{\begingroup \@sanitize@url \@href}%
\providecommand \@href[1]{\@@startlink{#1}\@@href}%
\providecommand \@@href[1]{\endgroup#1\@@endlink}%
\providecommand \@sanitize@url [0]{\catcode `\\12\catcode `\$12\catcode
  `\&12\catcode `\#12\catcode `\^12\catcode `\_12\catcode `\%12\relax}%
\providecommand \@@startlink[1]{}%
\providecommand \@@endlink[0]{}%
\providecommand \url  [0]{\begingroup\@sanitize@url \@url }%
\providecommand \@url [1]{\endgroup\@href {#1}{\urlprefix }}%
\providecommand \urlprefix  [0]{URL }%
\providecommand \Eprint [0]{\href }%
\providecommand \doibase [0]{http://dx.doi.org/}%
\providecommand \selectlanguage [0]{\@gobble}%
\providecommand \bibinfo  [0]{\@secondoftwo}%
\providecommand \bibfield  [0]{\@secondoftwo}%
\providecommand \translation [1]{[#1]}%
\providecommand \BibitemOpen [0]{}%
\providecommand \bibitemStop [0]{}%
\providecommand \bibitemNoStop [0]{.\EOS\space}%
\providecommand \EOS [0]{\spacefactor3000\relax}%
\providecommand \BibitemShut  [1]{\csname bibitem#1\endcsname}%
\let\auto@bib@innerbib\@empty
\bibitem [{\citenamefont {Gavela}\ \emph {et~al.}(1994)\citenamefont {Gavela},
  \citenamefont {Hernandez}, \citenamefont {Orloff},\ and\ \citenamefont
  {Pene}}]{Gavela:1993ts}%
  \BibitemOpen
  \bibfield  {author} {\bibinfo {author} {\bibfnamefont {M.~B.}\ \bibnamefont
  {Gavela}}, \bibinfo {author} {\bibfnamefont {P.}~\bibnamefont {Hernandez}},
  \bibinfo {author} {\bibfnamefont {J.}~\bibnamefont {Orloff}}, \ and\ \bibinfo
  {author} {\bibfnamefont {O.}~\bibnamefont {Pene}},\ }\href {\doibase
  10.1142/S0217732394000629} {\bibfield  {journal} {\bibinfo  {journal} {Mod.
  Phys. Lett.}\ }\textbf {\bibinfo {volume} {A9}},\ \bibinfo {pages} {795}
  (\bibinfo {year} {1994})},\ \Eprint {http://arxiv.org/abs/hep-ph/9312215
  [hep-ph]} {arXiv:hep-ph/9312215 [hep-ph]} \BibitemShut {NoStop}%
\bibitem [{\citenamefont {Huet}\ and\ \citenamefont
  {Sather}(1995)}]{Huet:1994jb}%
  \BibitemOpen
  \bibfield  {author} {\bibinfo {author} {\bibfnamefont {P.}~\bibnamefont
  {Huet}}\ and\ \bibinfo {author} {\bibfnamefont {E.}~\bibnamefont {Sather}},\
  }\href {\doibase 10.1103/PhysRevD.51.379} {\bibfield  {journal} {\bibinfo
  {journal} {Phys. Rev.}\ }\textbf {\bibinfo {volume} {D51}},\ \bibinfo {pages}
  {379} (\bibinfo {year} {1995})},\ \Eprint
  {http://arxiv.org/abs/hep-ph/9404302 [hep-ph]} {arXiv:hep-ph/9404302
  [hep-ph]} \BibitemShut {NoStop}%
\bibitem [{\citenamefont {Tanabashi}\ \emph {et~al.}(2018)\citenamefont
  {Tanabashi} \emph {et~al.}}]{Tanabashi:2018oca}%
  \BibitemOpen
  \bibfield  {author} {\bibinfo {author} {\bibfnamefont {M.}~\bibnamefont
  {Tanabashi}} \emph {et~al.} (\bibinfo {collaboration} {Particle Data
  Group}),\ }\href {\doibase 10.1103/PhysRevD.98.030001} {\bibfield  {journal}
  {\bibinfo  {journal} {Phys. Rev.}\ }\textbf {\bibinfo {volume} {D98}},\
  \bibinfo {pages} {030001} (\bibinfo {year} {2018})}\BibitemShut {NoStop}%
\bibitem [{\citenamefont {Cline}(2006)}]{Cline:2006ts}%
  \BibitemOpen
  \bibfield  {author} {\bibinfo {author} {\bibfnamefont {J.~M.}\ \bibnamefont
  {Cline}}\ }(\bibinfo {year} {2006})\ \Eprint
  {http://arxiv.org/abs/hep-ph/0609145} {arXiv:hep-ph/0609145 [hep-ph]}
  \BibitemShut {NoStop}%
\bibitem [{\citenamefont {Morrissey}\ and\ \citenamefont
  {Ramsey-Musolf}(2012)}]{Morrissey:2012db}%
  \BibitemOpen
  \bibfield  {author} {\bibinfo {author} {\bibfnamefont {D.~E.}\ \bibnamefont
  {Morrissey}}\ and\ \bibinfo {author} {\bibfnamefont {M.~J.}\ \bibnamefont
  {Ramsey-Musolf}},\ }\href {\doibase 10.1088/1367-2630/14/12/125003}
  {\bibfield  {journal} {\bibinfo  {journal} {New J. Phys.}\ }\textbf {\bibinfo
  {volume} {14}},\ \bibinfo {pages} {125003} (\bibinfo {year} {2012})},\
  \Eprint {http://arxiv.org/abs/1206.2942 [hep-ph]} {arXiv:1206.2942 [hep-ph]}
  \BibitemShut {NoStop}%
\bibitem [{\citenamefont {Konstandin}(2013)}]{Konstandin:2013caa}%
  \BibitemOpen
  \bibfield  {author} {\bibinfo {author} {\bibfnamefont {T.}~\bibnamefont
  {Konstandin}},\ }\href {\doibase 10.3367/UFNe.0183.201308a.0785} {\bibfield
  {journal} {\bibinfo  {journal} {Phys. Usp.}\ }\textbf {\bibinfo {volume}
  {56}},\ \bibinfo {pages} {747} (\bibinfo {year} {2013})},\ \Eprint
  {http://arxiv.org/abs/1302.6713} {arXiv:1302.6713 [hep-ph]} \BibitemShut
  {NoStop}%
\bibitem [{\citenamefont {Joyce}\ \emph {et~al.}(1994)\citenamefont {Joyce},
  \citenamefont {Prokopec},\ and\ \citenamefont {Turok}}]{Joyce:1994bi}%
  \BibitemOpen
  \bibfield  {author} {\bibinfo {author} {\bibfnamefont {M.}~\bibnamefont
  {Joyce}}, \bibinfo {author} {\bibfnamefont {T.}~\bibnamefont {Prokopec}}, \
  and\ \bibinfo {author} {\bibfnamefont {N.}~\bibnamefont {Turok}},\ }\href
  {\doibase 10.1016/0370-2693(94)91377-3} {\bibfield  {journal} {\bibinfo
  {journal} {Phys.\ Lett.\ B}\ }\textbf {\bibinfo {volume} {338}},\ \bibinfo
  {pages} {269} (\bibinfo {year} {1994})},\ \Eprint
  {http://arxiv.org/abs/hep-ph/9401352} {arXiv:hep-ph/9401352} \BibitemShut
  {NoStop}%
\bibitem [{\citenamefont {Cohen}\ \emph {et~al.}(1994)\citenamefont {Cohen},
  \citenamefont {Kaplan},\ and\ \citenamefont {Nelson}}]{Cohen:1994ss}%
  \BibitemOpen
  \bibfield  {author} {\bibinfo {author} {\bibfnamefont {A.~G.}\ \bibnamefont
  {Cohen}}, \bibinfo {author} {\bibfnamefont {D.}~\bibnamefont {Kaplan}}, \
  and\ \bibinfo {author} {\bibfnamefont {A.}~\bibnamefont {Nelson}},\ }\href
  {\doibase 10.1016/0370-2693(94)00935-X} {\bibfield  {journal} {\bibinfo
  {journal} {Phys.\ Lett.\ B}\ }\textbf {\bibinfo {volume} {336}},\ \bibinfo
  {pages} {41} (\bibinfo {year} {1994})},\ \Eprint
  {http://arxiv.org/abs/hep-ph/9406345} {arXiv:hep-ph/9406345} \BibitemShut
  {NoStop}%
\bibitem [{\citenamefont {Huet}\ and\ \citenamefont
  {Nelson}(1996)}]{Huet:1995sh}%
  \BibitemOpen
  \bibfield  {author} {\bibinfo {author} {\bibfnamefont {P.}~\bibnamefont
  {Huet}}\ and\ \bibinfo {author} {\bibfnamefont {A.~E.}\ \bibnamefont
  {Nelson}},\ }\href {\doibase 10.1103/PhysRevD.53.4578} {\bibfield  {journal}
  {\bibinfo  {journal} {Phys.\ Rev.\ D}\ }\textbf {\bibinfo {volume} {53}},\
  \bibinfo {pages} {4578} (\bibinfo {year} {1996})},\ \Eprint
  {http://arxiv.org/abs/hep-ph/9506477} {arXiv:hep-ph/9506477} \BibitemShut
  {NoStop}%
\bibitem [{\citenamefont {Riotto}(1998)}]{Riotto:1997vy}%
  \BibitemOpen
  \bibfield  {author} {\bibinfo {author} {\bibfnamefont {A.}~\bibnamefont
  {Riotto}},\ }\href {\doibase 10.1016/S0550-3213(98)00159-X} {\bibfield
  {journal} {\bibinfo  {journal} {Nucl.\ Phys.\ B}\ }\textbf {\bibinfo {volume}
  {518}},\ \bibinfo {pages} {339} (\bibinfo {year} {1998})},\ \Eprint
  {http://arxiv.org/abs/hep-ph/9712221} {arXiv:hep-ph/9712221} \BibitemShut
  {NoStop}%
\bibitem [{\citenamefont {Cline}\ \emph {et~al.}(2000)\citenamefont {Cline},
  \citenamefont {Joyce},\ and\ \citenamefont {Kainulainen}}]{Cline:2000nw}%
  \BibitemOpen
  \bibfield  {author} {\bibinfo {author} {\bibfnamefont {J.~M.}\ \bibnamefont
  {Cline}}, \bibinfo {author} {\bibfnamefont {M.}~\bibnamefont {Joyce}}, \ and\
  \bibinfo {author} {\bibfnamefont {K.}~\bibnamefont {Kainulainen}},\ }\href
  {\doibase 10.1088/1126-6708/2000/07/018} {\bibfield  {journal} {\bibinfo
  {journal} {JHEP}\ }\textbf {\bibinfo {volume} {07}},\ \bibinfo {pages} {018}
  (\bibinfo {year} {2000})},\ \Eprint {http://arxiv.org/abs/hep-ph/0006119}
  {arXiv:hep-ph/0006119} \BibitemShut {NoStop}%
\bibitem [{\citenamefont {Chung}\ \emph {et~al.}(2010)\citenamefont {Chung},
  \citenamefont {Garbrecht}, \citenamefont {Ramsey-Musolf},\ and\ \citenamefont
  {Tulin}}]{Chung:2009cb}%
  \BibitemOpen
  \bibfield  {author} {\bibinfo {author} {\bibfnamefont {D.~J.}\ \bibnamefont
  {Chung}}, \bibinfo {author} {\bibfnamefont {B.}~\bibnamefont {Garbrecht}},
  \bibinfo {author} {\bibfnamefont {M.~J.}\ \bibnamefont {Ramsey-Musolf}}, \
  and\ \bibinfo {author} {\bibfnamefont {S.}~\bibnamefont {Tulin}},\ }\href
  {\doibase 10.1103/PhysRevD.81.063506} {\bibfield  {journal} {\bibinfo
  {journal} {Phys. Rev. D}\ }\textbf {\bibinfo {volume} {81}},\ \bibinfo
  {pages} {063506} (\bibinfo {year} {2010})},\ \Eprint
  {http://arxiv.org/abs/0905.4509} {arXiv:0905.4509 [hep-ph]} \BibitemShut
  {NoStop}%
\bibitem [{\citenamefont {Guo}\ \emph {et~al.}(2017)\citenamefont {Guo},
  \citenamefont {Li}, \citenamefont {Liu}, \citenamefont {Ramsey-Musolf},\ and\
  \citenamefont {Shu}}]{Guo:2016ixx}%
  \BibitemOpen
  \bibfield  {author} {\bibinfo {author} {\bibfnamefont {H.-K.}\ \bibnamefont
  {Guo}}, \bibinfo {author} {\bibfnamefont {Y.-Y.}\ \bibnamefont {Li}},
  \bibinfo {author} {\bibfnamefont {T.}~\bibnamefont {Liu}}, \bibinfo {author}
  {\bibfnamefont {M.}~\bibnamefont {Ramsey-Musolf}}, \ and\ \bibinfo {author}
  {\bibfnamefont {J.}~\bibnamefont {Shu}},\ }\href {\doibase
  10.1103/PhysRevD.96.115034} {\bibfield  {journal} {\bibinfo  {journal} {Phys.
  Rev.}\ }\textbf {\bibinfo {volume} {D96}},\ \bibinfo {pages} {115034}
  (\bibinfo {year} {2017})},\ \Eprint {http://arxiv.org/abs/1609.09849
  [hep-ph]} {arXiv:1609.09849 [hep-ph]} \BibitemShut {NoStop}%
\bibitem [{\citenamefont {de~Vries}\ \emph {et~al.}(2019)\citenamefont
  {de~Vries}, \citenamefont {Postma},\ and\ \citenamefont {van~de
  Vis}}]{deVries:2018tgs}%
  \BibitemOpen
  \bibfield  {author} {\bibinfo {author} {\bibfnamefont {J.}~\bibnamefont
  {de~Vries}}, \bibinfo {author} {\bibfnamefont {M.}~\bibnamefont {Postma}}, \
  and\ \bibinfo {author} {\bibfnamefont {J.}~\bibnamefont {van~de Vis}},\
  }\href {\doibase 10.1007/JHEP04(2019)024} {\bibfield  {journal} {\bibinfo
  {journal} {JHEP}\ }\textbf {\bibinfo {volume} {04}},\ \bibinfo {pages} {024}
  (\bibinfo {year} {2019})},\ \Eprint {http://arxiv.org/abs/1811.11104
  [hep-ph]} {arXiv:1811.11104 [hep-ph]} \BibitemShut {NoStop}%
\bibitem [{\citenamefont {Joyce}\ \emph {et~al.}(1996)\citenamefont {Joyce},
  \citenamefont {Prokopec},\ and\ \citenamefont {Turok}}]{Joyce_1996}%
  \BibitemOpen
  \bibfield  {author} {\bibinfo {author} {\bibfnamefont {M.}~\bibnamefont
  {Joyce}}, \bibinfo {author} {\bibfnamefont {T.}~\bibnamefont {Prokopec}}, \
  and\ \bibinfo {author} {\bibfnamefont {N.}~\bibnamefont {Turok}},\ }\href
  {\doibase 10.1103/physrevd.53.2930} {\bibfield  {journal} {\bibinfo
  {journal} {Phys. Rev. D}\ }\textbf {\bibinfo {volume} {53}},\ \bibinfo
  {pages} {2930} (\bibinfo {year} {1996})}\BibitemShut {NoStop}%
\bibitem [{\citenamefont {Fuchs}\ \emph
  {et~al.}(2020{\natexlab{a}})\citenamefont {Fuchs}, \citenamefont {Losada},
  \citenamefont {Nir},\ and\ \citenamefont {Viernik}}]{Fuchs:2020uoc}%
  \BibitemOpen
  \bibfield  {author} {\bibinfo {author} {\bibfnamefont {E.}~\bibnamefont
  {Fuchs}}, \bibinfo {author} {\bibfnamefont {M.}~\bibnamefont {Losada}},
  \bibinfo {author} {\bibfnamefont {Y.}~\bibnamefont {Nir}}, \ and\ \bibinfo
  {author} {\bibfnamefont {Y.}~\bibnamefont {Viernik}},\ }\href {\doibase
  10.1007/JHEP05(2020)056} {\bibfield  {journal} {\bibinfo  {journal} {JHEP}\
  }\textbf {\bibinfo {volume} {05}},\ \bibinfo {pages} {056} (\bibinfo {year}
  {2020}{\natexlab{a}})},\ \Eprint {http://arxiv.org/abs/2003.00099}
  {arXiv:2003.00099 [hep-ph]} \BibitemShut {NoStop}%
\bibitem [{\citenamefont {White}(2016)}]{White:2015bva}%
  \BibitemOpen
  \bibfield  {author} {\bibinfo {author} {\bibfnamefont {G.~A.}\ \bibnamefont
  {White}},\ }\href {\doibase 10.1103/PhysRevD.93.043504} {\bibfield  {journal}
  {\bibinfo  {journal} {Phys. Rev.}\ }\textbf {\bibinfo {volume} {D93}},\
  \bibinfo {pages} {043504} (\bibinfo {year} {2016})},\ \Eprint
  {http://arxiv.org/abs/1510.03901 [hep-ph]} {arXiv:1510.03901 [hep-ph]}
  \BibitemShut {NoStop}%
\bibitem [{\citenamefont {Carena}\ \emph {et~al.}(2003)\citenamefont {Carena},
  \citenamefont {Quiros}, \citenamefont {Seco},\ and\ \citenamefont
  {Wagner}}]{Carena:2002ss}%
  \BibitemOpen
  \bibfield  {author} {\bibinfo {author} {\bibfnamefont {M.}~\bibnamefont
  {Carena}}, \bibinfo {author} {\bibfnamefont {M.}~\bibnamefont {Quiros}},
  \bibinfo {author} {\bibfnamefont {M.}~\bibnamefont {Seco}}, \ and\ \bibinfo
  {author} {\bibfnamefont {C.}~\bibnamefont {Wagner}},\ }\href {\doibase
  10.1016/S0550-3213(02)01065-9} {\bibfield  {journal} {\bibinfo  {journal}
  {Nucl. Phys. B}\ }\textbf {\bibinfo {volume} {650}},\ \bibinfo {pages} {24}
  (\bibinfo {year} {2003})},\ \Eprint {http://arxiv.org/abs/hep-ph/0208043}
  {arXiv:hep-ph/0208043} \BibitemShut {NoStop}%
\bibitem [{\citenamefont {Cirigliano}\ \emph {et~al.}(2006)\citenamefont
  {Cirigliano}, \citenamefont {Ramsey-Musolf}, \citenamefont {Tulin},\ and\
  \citenamefont {Lee}}]{Cirigliano_2006}%
  \BibitemOpen
  \bibfield  {author} {\bibinfo {author} {\bibfnamefont {V.}~\bibnamefont
  {Cirigliano}}, \bibinfo {author} {\bibfnamefont {M.~J.}\ \bibnamefont
  {Ramsey-Musolf}}, \bibinfo {author} {\bibfnamefont {S.}~\bibnamefont
  {Tulin}}, \ and\ \bibinfo {author} {\bibfnamefont {C.}~\bibnamefont {Lee}},\
  }\href {\doibase 10.1103/physrevd.73.115009} {\bibfield  {journal} {\bibinfo
  {journal} {Phys. Rev. D}\ }\textbf {\bibinfo {volume} {73}} (\bibinfo {year}
  {2006}),\ 10.1103/physrevd.73.115009}\BibitemShut {NoStop}%
\bibitem [{\citenamefont {Lee}\ \emph {et~al.}(2005)\citenamefont {Lee},
  \citenamefont {Cirigliano},\ and\ \citenamefont
  {Ramsey-Musolf}}]{Lee:2004we}%
  \BibitemOpen
  \bibfield  {author} {\bibinfo {author} {\bibfnamefont {C.}~\bibnamefont
  {Lee}}, \bibinfo {author} {\bibfnamefont {V.}~\bibnamefont {Cirigliano}}, \
  and\ \bibinfo {author} {\bibfnamefont {M.~J.}\ \bibnamefont
  {Ramsey-Musolf}},\ }\href {\doibase 10.1103/PhysRevD.71.075010} {\bibfield
  {journal} {\bibinfo  {journal} {Phys. Rev.}\ }\textbf {\bibinfo {volume}
  {D71}},\ \bibinfo {pages} {075010} (\bibinfo {year} {2005})},\ \Eprint
  {http://arxiv.org/abs/hep-ph/0412354 [hep-ph]} {arXiv:hep-ph/0412354
  [hep-ph]} \BibitemShut {NoStop}%
\bibitem [{\citenamefont {de~Vries}\ \emph {et~al.}(2018)\citenamefont
  {de~Vries}, \citenamefont {Postma}, \citenamefont {van~de Vis},\ and\
  \citenamefont {White}}]{deVries:2017ncy}%
  \BibitemOpen
  \bibfield  {author} {\bibinfo {author} {\bibfnamefont {J.}~\bibnamefont
  {de~Vries}}, \bibinfo {author} {\bibfnamefont {M.}~\bibnamefont {Postma}},
  \bibinfo {author} {\bibfnamefont {J.}~\bibnamefont {van~de Vis}}, \ and\
  \bibinfo {author} {\bibfnamefont {G.}~\bibnamefont {White}},\ }\href
  {\doibase 10.1007/JHEP01(2018)089} {\bibfield  {journal} {\bibinfo  {journal}
  {JHEP}\ }\textbf {\bibinfo {volume} {01}},\ \bibinfo {pages} {089} (\bibinfo
  {year} {2018})},\ \Eprint {http://arxiv.org/abs/1710.04061 [hep-ph]}
  {arXiv:1710.04061 [hep-ph]} \BibitemShut {NoStop}%
\bibitem [{\citenamefont {Trodden}(1999)}]{Trodden:1998ym}%
  \BibitemOpen
  \bibfield  {author} {\bibinfo {author} {\bibfnamefont {M.}~\bibnamefont
  {Trodden}},\ }\href {\doibase 10.1103/RevModPhys.71.1463} {\bibfield
  {journal} {\bibinfo  {journal} {Rev.\ Mod.\ Phys.}\ }\textbf {\bibinfo
  {volume} {71}},\ \bibinfo {pages} {1463} (\bibinfo {year} {1999})},\ \Eprint
  {http://arxiv.org/abs/hep-ph/9803479} {arXiv:hep-ph/9803479} \BibitemShut
  {NoStop}%
\bibitem [{\citenamefont {Fuchs}\ \emph
  {et~al.}(2020{\natexlab{b}})\citenamefont {Fuchs}, \citenamefont {Losada},
  \citenamefont {Nir},\ and\ \citenamefont {Viernik}}]{Fuchs:2019ore}%
  \BibitemOpen
  \bibfield  {author} {\bibinfo {author} {\bibfnamefont {E.}~\bibnamefont
  {Fuchs}}, \bibinfo {author} {\bibfnamefont {M.}~\bibnamefont {Losada}},
  \bibinfo {author} {\bibfnamefont {Y.}~\bibnamefont {Nir}}, \ and\ \bibinfo
  {author} {\bibfnamefont {Y.}~\bibnamefont {Viernik}},\ }\href {\doibase
  10.1103/PhysRevLett.124.181801} {\bibfield  {journal} {\bibinfo  {journal}
  {Phys. Rev. Lett.}\ }\textbf {\bibinfo {volume} {124}},\ \bibinfo {pages}
  {181801} (\bibinfo {year} {2020}{\natexlab{b}})},\ \Eprint
  {http://arxiv.org/abs/1911.08495} {arXiv:1911.08495 [hep-ph]} \BibitemShut
  {NoStop}%
\bibitem [{\citenamefont {Postma}\ and\ \citenamefont {Van
  De~Vis}(2020)}]{Postma:2019scv}%
  \BibitemOpen
  \bibfield  {author} {\bibinfo {author} {\bibfnamefont {M.}~\bibnamefont
  {Postma}}\ and\ \bibinfo {author} {\bibfnamefont {J.}~\bibnamefont {Van
  De~Vis}},\ }\href {\doibase 10.1007/JHEP02(2020)090} {\bibfield  {journal}
  {\bibinfo  {journal} {JHEP}\ }\textbf {\bibinfo {volume} {02}},\ \bibinfo
  {pages} {090} (\bibinfo {year} {2020})},\ \Eprint
  {http://arxiv.org/abs/1910.11794} {arXiv:1910.11794 [hep-ph]} \BibitemShut
  {NoStop}%
\bibitem [{\citenamefont {Moore}(1997)}]{Moore:1997im}%
  \BibitemOpen
  \bibfield  {author} {\bibinfo {author} {\bibfnamefont {G.~D.}\ \bibnamefont
  {Moore}},\ }\href {\doibase 10.1016/S0370-2693(97)01046-0} {\bibfield
  {journal} {\bibinfo  {journal} {Phys. Lett. B}\ }\textbf {\bibinfo {volume}
  {412}},\ \bibinfo {pages} {359} (\bibinfo {year} {1997})},\ \Eprint
  {http://arxiv.org/abs/hep-ph/9705248} {arXiv:hep-ph/9705248} \BibitemShut
  {NoStop}%
\bibitem [{\citenamefont {Moore}(2000)}]{Moore:2000mx}%
  \BibitemOpen
  \bibfield  {author} {\bibinfo {author} {\bibfnamefont {G.~D.}\ \bibnamefont
  {Moore}},\ }\href {\doibase 10.1103/PhysRevD.62.085011} {\bibfield  {journal}
  {\bibinfo  {journal} {Phys. Rev. D}\ }\textbf {\bibinfo {volume} {62}},\
  \bibinfo {pages} {085011} (\bibinfo {year} {2000})},\ \Eprint
  {http://arxiv.org/abs/hep-ph/0001216} {arXiv:hep-ph/0001216} \BibitemShut
  {NoStop}%
\bibitem [{\citenamefont {D'Onofrio}\ \emph {et~al.}(2014)\citenamefont
  {D'Onofrio}, \citenamefont {Rummukainen},\ and\ \citenamefont
  {Tranberg}}]{DOnofrio:2014rug}%
  \BibitemOpen
  \bibfield  {author} {\bibinfo {author} {\bibfnamefont {M.}~\bibnamefont
  {D'Onofrio}}, \bibinfo {author} {\bibfnamefont {K.}~\bibnamefont
  {Rummukainen}}, \ and\ \bibinfo {author} {\bibfnamefont {A.}~\bibnamefont
  {Tranberg}},\ }\href {\doibase 10.1103/PhysRevLett.113.141602} {\bibfield
  {journal} {\bibinfo  {journal} {Phys. Rev. Lett.}\ }\textbf {\bibinfo
  {volume} {113}},\ \bibinfo {pages} {141602} (\bibinfo {year} {2014})},\
  \Eprint {http://arxiv.org/abs/1404.3565} {arXiv:1404.3565 [hep-ph]}
  \BibitemShut {NoStop}%
\bibitem [{\citenamefont {Kozaczuk}(2015)}]{Kozaczuk:2015owa}%
  \BibitemOpen
  \bibfield  {author} {\bibinfo {author} {\bibfnamefont {J.}~\bibnamefont
  {Kozaczuk}},\ }\href {\doibase 10.1007/JHEP10(2015)135} {\bibfield  {journal}
  {\bibinfo  {journal} {JHEP}\ }\textbf {\bibinfo {volume} {10}},\ \bibinfo
  {pages} {135} (\bibinfo {year} {2015})},\ \Eprint
  {http://arxiv.org/abs/1506.04741} {arXiv:1506.04741 [hep-ph]} \BibitemShut
  {NoStop}%
\bibitem [{\citenamefont {Cline}\ and\ \citenamefont
  {Kainulainen}(2020)}]{Cline:2020jre}%
  \BibitemOpen
  \bibfield  {author} {\bibinfo {author} {\bibfnamefont {J.~M.}\ \bibnamefont
  {Cline}}\ and\ \bibinfo {author} {\bibfnamefont {K.}~\bibnamefont
  {Kainulainen}},\ }\href {\doibase 10.1103/PhysRevD.101.063525} {\bibfield
  {journal} {\bibinfo  {journal} {Phys. Rev. D}\ }\textbf {\bibinfo {volume}
  {101}},\ \bibinfo {pages} {063525} (\bibinfo {year} {2020})},\ \Eprint
  {http://arxiv.org/abs/2001.00568} {arXiv:2001.00568 [hep-ph]} \BibitemShut
  {NoStop}%
\bibitem [{\citenamefont {Enqvist}\ \emph {et~al.}(1998)\citenamefont
  {Enqvist}, \citenamefont {Riotto},\ and\ \citenamefont
  {Vilja}}]{Enqvist_1998}%
  \BibitemOpen
  \bibfield  {author} {\bibinfo {author} {\bibfnamefont {K.}~\bibnamefont
  {Enqvist}}, \bibinfo {author} {\bibfnamefont {A.}~\bibnamefont {Riotto}}, \
  and\ \bibinfo {author} {\bibfnamefont {I.}~\bibnamefont {Vilja}},\ }\href
  {\doibase 10.1016/s0370-2693(98)00963-0} {\bibfield  {journal} {\bibinfo
  {journal} {Phys. Lett. B}\ }\textbf {\bibinfo {volume} {438}},\ \bibinfo
  {pages} {273} (\bibinfo {year} {1998})}\BibitemShut {NoStop}%
\bibitem [{\citenamefont {Elmfors}\ \emph {et~al.}(1999)\citenamefont
  {Elmfors}, \citenamefont {Enqvist}, \citenamefont {Riotto},\ and\
  \citenamefont {Vilja}}]{Elmfors_1999}%
  \BibitemOpen
  \bibfield  {author} {\bibinfo {author} {\bibfnamefont {P.}~\bibnamefont
  {Elmfors}}, \bibinfo {author} {\bibfnamefont {K.}~\bibnamefont {Enqvist}},
  \bibinfo {author} {\bibfnamefont {A.}~\bibnamefont {Riotto}}, \ and\ \bibinfo
  {author} {\bibfnamefont {I.}~\bibnamefont {Vilja}},\ }\href {\doibase
  10.1016/s0370-2693(99)00169-0} {\bibfield  {journal} {\bibinfo  {journal}
  {Phys. Lett. B}\ }\textbf {\bibinfo {volume} {452}},\ \bibinfo {pages} {279}
  (\bibinfo {year} {1999})}\BibitemShut {NoStop}%
\bibitem [{\citenamefont {Bodeker}\ \emph {et~al.}(2000)\citenamefont
  {Bodeker}, \citenamefont {Moore},\ and\ \citenamefont
  {Rummukainen}}]{Bodeker:1999gx}%
  \BibitemOpen
  \bibfield  {author} {\bibinfo {author} {\bibfnamefont {D.}~\bibnamefont
  {Bodeker}}, \bibinfo {author} {\bibfnamefont {G.~D.}\ \bibnamefont {Moore}},
  \ and\ \bibinfo {author} {\bibfnamefont {K.}~\bibnamefont {Rummukainen}},\
  }\href {\doibase 10.1103/PhysRevD.61.056003} {\bibfield  {journal} {\bibinfo
  {journal} {Phys. Rev.}\ }\textbf {\bibinfo {volume} {D61}},\ \bibinfo {pages}
  {056003} (\bibinfo {year} {2000})},\ \Eprint
  {http://arxiv.org/abs/hep-ph/9907545 [hep-ph]} {arXiv:hep-ph/9907545
  [hep-ph]} \BibitemShut {NoStop}%
\bibitem [{\citenamefont {Moore}\ and\ \citenamefont
  {Tassler}(2011)}]{Moore:2010jd}%
  \BibitemOpen
  \bibfield  {author} {\bibinfo {author} {\bibfnamefont {G.~D.}\ \bibnamefont
  {Moore}}\ and\ \bibinfo {author} {\bibfnamefont {M.}~\bibnamefont
  {Tassler}},\ }\href {\doibase 10.1007/JHEP02(2011)105} {\bibfield  {journal}
  {\bibinfo  {journal} {JHEP}\ }\textbf {\bibinfo {volume} {02}},\ \bibinfo
  {pages} {105} (\bibinfo {year} {2011})},\ \Eprint
  {http://arxiv.org/abs/1011.1167 [hep-ph]} {arXiv:1011.1167 [hep-ph]}
  \BibitemShut {NoStop}%
\end{thebibliography}%
\end{document}